\documentclass[twocolumn]{aastex62}

\usepackage{graphicx}
\usepackage{xcolor}
\usepackage{natbib}
\usepackage{amsmath}
\usepackage[subfigure]{tocloft}
\usepackage{subfigure}
\usepackage{booktabs}
\usepackage{threeparttable}
\usepackage{CJK}
\usepackage{url}
\usepackage{longtable}

\newcounter{RomanNumber}

\newcommand{\degree}{^\circ}
\newcommand{\ue}{\mathrm{e}}

\received{xxx}
\revised{xxx}
\accepted{xxx}

\graphicspath{{./}{fig/}}

\shorttitle{AT 2019avd}

\begin{document}
\begin{CJK*}{UTF8}{gbsn}

\title{\large{\bf{AT 2019avd: A tidal disruption event with a two-phase evolution}}}

\author{Jin-Hong Chen (陈劲鸿)}
\affiliation{School of Physics and Astronomy, Sun Yat-Sen University, Zhuhai, 519000, China}
\author{Li-Ming Dou (窦立明)}
\affiliation{Center for Astrophysics, Guangzhou University, Guangzhou, 510006, China}
\affiliation{Astronomy Science and Technology Research Laboratory of Department of Education of Guangdong Province, Guangzhou, 510006, China}
\author{Rong-Feng Shen (申荣锋)}
\affiliation{School of Physics and Astronomy, Sun Yat-Sen University, Zhuhai, 519000, China}

\email{chenjh258@mail2.sysu.edu.cn, doulm@gzhu.edu.cn, \\ 
shenrf3@mail.sysu.edu.cn}

\begin{abstract}
Tidal disruption events (TDEs) can uncover the quiescent supermassive black holes (SMBHs) at the center of galaxies. After the disruption of a star by a SMBH, the highly elliptical orbit of the debris stream will be gradually circularized due to the self-crossing, and then the circularized debris will form an accretion disk. The recent TDE candidate AT 2019avd has double peaks in its optical light curve, and the X-ray emerges near the second peak. The durations of the peaks are $\sim 400$ and $600$ days, respectively, and the separation between them is $\sim 700$ days. We fit its spectral energy distribution (SED) and analyze its light curves in the optical/UV, mid-infrared, and X-ray bands. We find that this source can be interpreted as a two-phase scenario in which the first phase is dominated by the stream circularization, and the second phase is the delayed accretion. We use the succession of the self-crossing model and the delayed accretion model to fit the first and the second peaks, respectively. The fitting result implies that AT 2019avd can be interpreted by the partial disruption of a $0.9\ M_{\odot}$ star by a $7 \times 10^6\ M_{\odot}$ SMBH, but this result is sensitive to the stellar model. Furthermore, we find the large-amplitude (by factors up to $\sim 5$) X-ray variability in AT 2019avd can be interpreted as the rigid-body precession of the misaligned disk due to the Lense--Thirring effect of a spinning SMBH, with the precession period of $10 - 25$ days.
\end{abstract}

\keywords{accretion, accretion disks - black hole physics - galaxies: nuclei - tidal disruption}

\section{Introduction} \label{sec_intro}
Occasionally, a star is disrupted by the supermassive black hole (SMBH) in the center of a galaxy, and then its debris will experience a circularization process to form an accretion disk \citep{Rees_Tidal_1988,Hayasaki_Finite_2013,Dai_Soft_2015,Bonnerot_Disc_2016,Bonnerot_Long_2017}. In such a so-called tidal disruption event (TDE), the circularization process is mainly driven by the self-collision of the debris stream near the apocenter. The early optical/UV emission might come from the circularization process \citep{Jiang_Prompt_2016,Lu_Self_2020}. During or after the circularization, some debris should be close to and be accreted by the central SMBH. Then the accretion process near the SMBH can emit significant amount of X-ray photons. 

The details of the circularization process and the issue of the disk formation have long been debated. Most of the simulations indicate that the circularization process may be inefficient if the pericenter of the star is not so close to the SMBH \citep{Hayasaki_Finite_2013,Bonnerot_Disc_2016,Bonnerot_Long_2017}; thus, the formation of accretion disk would be delayed. The inefficient circularization process will lead to two peaks in the light curve \citep{Chen_Light_2021}, especially for the partial TDEs in which the pericenter of the tidally disrupted star is slightly farther away than the tidal radius. The first and the second peaks correspond to the circularization and the accretion processes, respectively. Here we call it as the two-phase model.

It is commonly believed that during the circularization process, the optical/UV emission originates from the self-collision near the apocenter as the stretched stream shocks itself near the apocenter due to the apsidal precession \citep{Jiang_Prompt_2016,Lu_Self_2020}. During the accretion process, the optical/UV emission may come from an outflow when the accretion is super-Eddington \citep{Strubbe_Optical_2009, Lodato_Multiband_2011,Metzger_A_2016}.

If the early optical/UV emission comes from the circularization process, the size of the optical/UV emitting region would be close to the self-collision radius, as expected in the theory \citep{Piran_Disk_2015}, which is consistent with the observations \citep{Wevers_Evidence_2019}. 

Firmer evidence of the two-phase model is from a study of the late-time UV observations of TDEs by \cite{van_Velzen_Late_2019}. They found that the late-time emission, which is different from that in the early time, can be explained by the ongoing disk accretion, which causes the light curve to show a late-time flattening or peak. It implies that the optical/UV component of the early emission is not powered by accretion; instead, it can be reasonably explained by the circularization model.

A recently reported special TDE candidate, AT 2019avd \citep{Malyali_AT2019avd_2021}, whose optical light curve has two peaks with the X-ray emerging near the second peak, is consistent with this two-phase picture. In particular, the presence of Bowen fluorescence near the second peak, which is triggered by the high-energy photons \citep{Bowen_The_1928}, is a signature of an activated and ongoing disk accretion.

In Section \ref{sec_obs}, we present the observations of AT 2019avd, then fit and analyze its spectral energy distribution (SED) in the optical/UV, mid-infrared, and X-ray in Section \ref{sec:spec}. In Sections \ref{sec_cir} and \ref{sec_acc}, we present the model of the circularization and accretion process. In Section \ref{sec_fitting}, we fit the observational data using the model. In Section \ref{sec_xray_var}, we interpret the X-ray variability as the rigid disk precession. We summarize and discuss the results in Section \ref{sec_conclus}.

\section{OBSERVATIONS} \label{sec_obs}
\subsection{Optical and X-Ray discoveries} \label{subsec_discovery}

\cite{Malyali_AT2019avd_2021} reported AT 2019avd (ZTF19aaiqmgl) as a unique TDE candidate based on its TDE-like X-ray behavior. However, its double-peak optical light curve is unlike the typical TDEs. In this paper, we study and show that this transient belongs to a TDE whose early emission and the rebrightening comes from the circularization process of the debris and the delayed accretion process, respectively.

Nuclear transient AT 2019avd is in the inactive galaxy 2MASX J08233674+0423027 at $z = 0.029$ (corresponding to a luminosity distance of 130 Mpc adopting a flat $\Lambda$CDM cosmology), discovered by the Zwicky Transient Facility (ZTF) on 2019 February 9 \cite[MJD 58,523.2; ][]{Nordin_ZTF_2019}. We downloaded the ZTF optical light curve of AT 2019avd using the \textit{Lasair alert broker} \citep{Smith_Lasair_2019} \footnote{\url{https://lasair.roe.ac.uk/object/ZTF19aaiqmgl/}} and show them in the middle panel of Fig. \ref{fig:LC}. 

The first X-ray detection of AT 2019avd on 2020 April 28 (MJD 58,967.7) is from a dedicated search for candidate TDEs in the first eROSITA all-sky survey \citep{Malyali_AT2019avd_2021}. A series of follow-up observations were performed with the Neil Gehrels Swift Observatory.

Between the optical discovery and the eROSITA observation of AT 2019avd, there is no X-ray observation. The nondetection of Bowen fluorescence around the spectroscopic observation of the 2.56 m Nordic Optical Telescope (NOT) on 2019 March 15 (MJD 58,557) implies that the X-ray flux is low during the first optical peak, until the appearance of a Bowen feature around the observation of the Wide Field Spectrograph (WiFeS) on 2020 May 29 \cite[MJD 58,998; ][]{Malyali_AT2019avd_2021}. We infer from this fact that the accretion disk forms near the second optical peak.  

\cite{Frederick_A_2021} estimated the masses of the center SMBH as $\sim 10^6 M_{\odot}$ and $\sim 10^7 M_{\odot}$, which are derived through the virial method based on the FWHM H$\beta$ emission and the host galaxy luminosity, respectively.

\subsection{Swift follow-up} \label{subsec_swift_follow}
We downloaded and reduced the X-ray data \citep[Swift XRT;][]{Burrows_XRT_2005} with the HEASoft V6.26 package and the latest updated calibration files. We only use the observations in photon-counting mode. We extract the source spectra from the circular region with a radius of $47^{\prime \prime}.2$ and the background spectra from a large source-free annulus at the same source position center. 

In each of the observations, the X-ray spectra show low count rates in the hard band $>2$ keV; thus, we only show the $0.3-2$ keV count rate and present in the bottom panel of Fig. \ref{fig:LC}. The XRT 0.3 - 2 keV count rate increases from $\sim 0.03$ to $\sim 0.2\ {\rm counts\ s^{-1}}$ in $90$ days and evolves with drastic time variability afterward (by factors up to $\sim 5$). This X-ray flare lasts for about $1$ year, until its count rate drops to $\sim 10^{-3}\ {\rm counts\ s^{-1}}$.

The Ultraviolet and Optical Telescope \citep[Swift UVOT; ][]{Roming_UVOT_2005} observed the source with multiwavelength filters (\textit{V}, \textit{B}, \textit{U}, \textit{UVM2}, \textit{UVW1}, \textit{UVW2}) simultaneously with the XRT observations. The UVOT flux was extracted with the 'uvotsource' task from the circular region with a radius of $9^{\prime \prime}$ and the background flux from the nearby circular region with a $15^{\prime \prime}$ radius. 

The Galactic extinction in the optical band is obtained from NED \footnote{\url{https://ned.ipac.caltech.edu/forms/calculator.html}} and extend to the extinction curve model of \cite{Cardelli_Extinction_1989}. After correcting for the Galactic extinction, we subtract the host galaxy contribution with the same SED model of \cite{Malyali_AT2019avd_2021}. 

We list the UV photometry data in Table \ref{tab:uv} and present the host-subtracted photometry results in the \textit{UVW1}, \textit{UVW2}, \textit{UVM2} bands in the middle panel of Fig. \ref{fig:LC}. Overall, the UV light curve has the same evolution as the light curve in the \textit{g}, \textit{r} bands; therefore, the optical and the UV emissions might come from the same process or location.

\subsection{Mid-infrared} \label{subsec_mid}
\begin{figure}
\centering
 \includegraphics[scale=0.5]{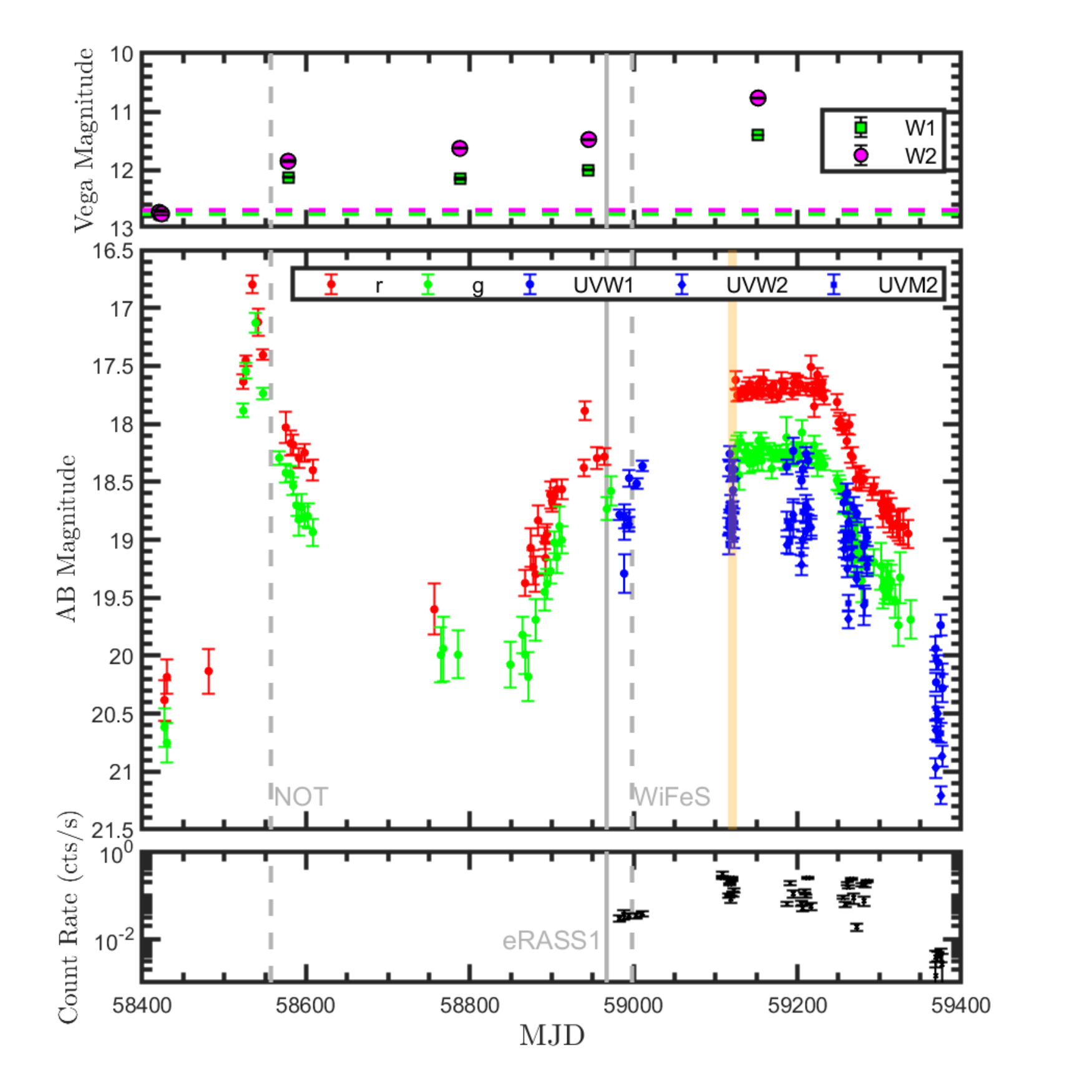}
\caption{NEOWISE-R (non-host subtracted, top) and ZTF and UVOT light curves of AT 2019avd (middle), with the XRT $0.3-2$ keV count rate history shown in the bottom panel. The ZTF and UVOT data are host-subtracted, and the UVOT data are corrected by the Galactic extinction. The NEOWISE-R observations preoutburst are observed with mean W1 and W2 shown in the top panel by the green and magenta dashed lines, respectively. The solid gray vertical line indicates the MJD of the eRASS1 observation, which is the first observation of AT 2019avd in the X-ray. The spectroscopic observations with the NOT on 2019 March15 (MJD 58,557) and WiFeS on 2020 May 29 \citep[MJD 58,998; ][]{Malyali_AT2019avd_2021} are marked by the dashed gray vertical lines. The trigger of Bowen fluorescence requires a high flux of far-UV or soft X-ray photons; therefore the appearance of a Bowen feature around the WiFeS observation might implicate the formation of an accretion disk near the second peak of the light curve. We use the nearly simultaneous observations shown in the orange shaded area to fit the optical/UV spectrum, which is shown in Fig. \ref{fig:opt_spec}. This figure is similar to Figure 5 in \cite{Malyali_AT2019avd_2021}, except that we update the observations of WISE, ZTF, and XRT.}
\label{fig:LC}
\end{figure}

In the top panel of Fig. \ref{fig:LC}, we also update and plot the mid-infrared data of AT 2019avd observed by the Wide-Field Infrared Survey Explorer (WISE) mission \citep{Wright_WISE_2010} and Near-Earth Object WISE \citep[NEOWISE-R; ][]{Mainzer_NEOWISE_2014}. We extract a total of $15$ epochs of mid-infrared photometry from the WISE/NEOWISE archive. There are 10-20 exposures in each epoch. We use only the best-quality single-frame images by selecting only detections with the data quality flag ``qual\_frame'' $> 0$. This leaves 9-17 measurements for each epoch. We then average these fluxes to obtain a mean value at each epoch. 

The W1- and W2-band light curves are shown in Fig \ref{fig:LC}. The mid-infrared is in the quiescent state before the first optical outburst, then follows the optical light curve to rise,;and it has a second rise when the optical light curve reaches its second peak. Therefore, the mid-infrared signature is associated with AT 2019avd. 

\section{Spectral Analysis}  \label{sec:spec}
\subsection{UV/Optical}        \label{subsec:spec_uvo}
\begin{figure}
\centering
 \includegraphics[scale=0.5]{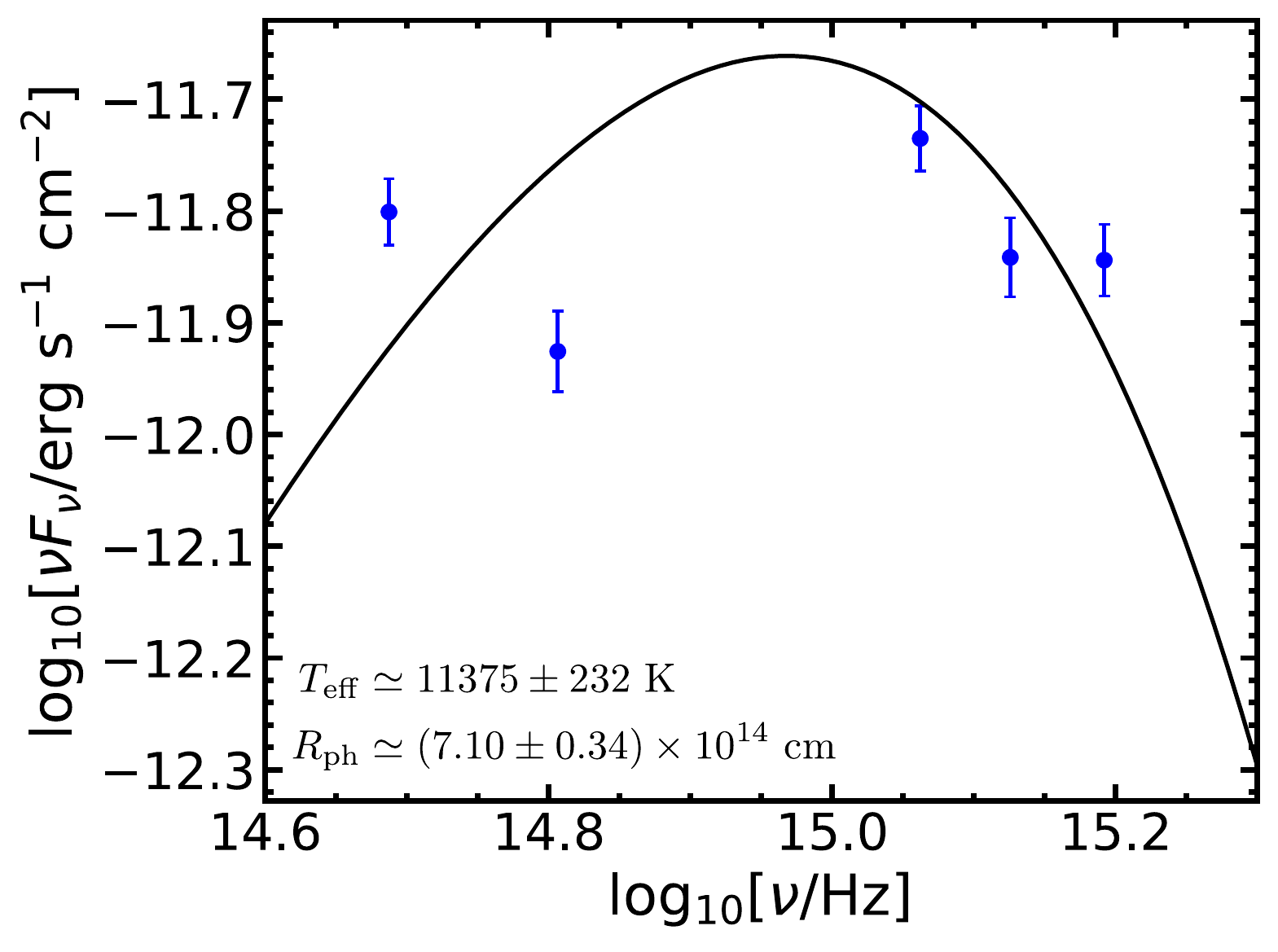}
\caption{The SED of AT 2019avd during nearly simultaneous ZTF optical and UVOT ultraviolet observations (with the host galaxy flux removed) from MJD 59,123 to MJD 59,124. We fit the spectra with a blackbody. The best-fit parameters are in the lower left corner of the figure with a $1\sigma$ error.}
\label{fig:opt_spec}
\end{figure}

We fit the optical/UV SED with the blackbody model using the least-squares method. We use the nearly simultaneous ZTF optical and UVOT ultraviolet observations from MJD 59,123 to MJD 59,124 to fit the SED. The effective temperature and the photospheric radius given by the fitting are $T_{\rm eff} \simeq 11,375 \pm 232\ {\rm K}$ and $R_{\rm ph} \simeq (7.10 \pm 0.34) \times 10^{14}\ {\rm cm}$, respectively. We plot the optical/UV SED in Fig. \ref{fig:opt_spec}.

\begin{figure}
\centering
 \includegraphics[scale=0.5]{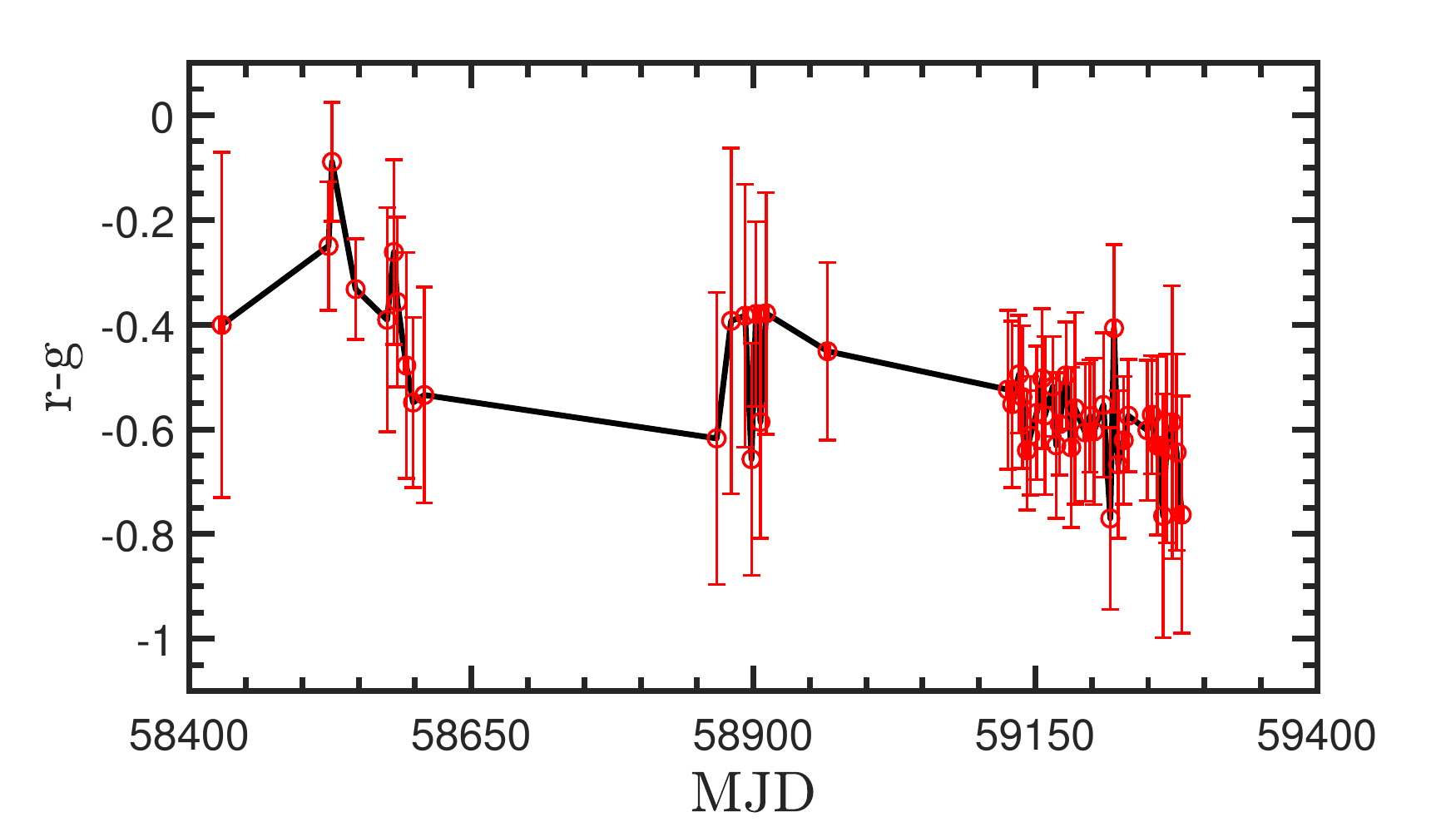}
\caption{The r--g colour evolution history of AT 2019avd. Here r--g is close to a constant with r--g $\simeq -0.53$ during the outburst. Therefore, it is reasonable to assume a constant blackbody temperature during the outburst.}
\label{fig:r_g}
\end{figure}

In order to obtain the luminosity history, we need to know the evolution of the SED. Unfortunately, we do not have enough UV observational data to derive the temperature evolution. Instead, we assume a constant effective temperature $T_{\rm eff} \simeq 11,375\ {\rm K}$ in the optical/UV band during the whole evolution. We need to verify the validity of the constant temperature assumption here. In Fig. \ref{fig:r_g}, we plot the r--g color evolution history, which has only a very slight drop with an average value r--g $ \simeq -0.53$. Therefore, it is reasonable to assume a constant blackbody temperature during the outburst.


\subsection{Mid-infrared}        \label{subsec:spec_wise}
\begin{figure}
\centering
 \includegraphics[scale=0.5]{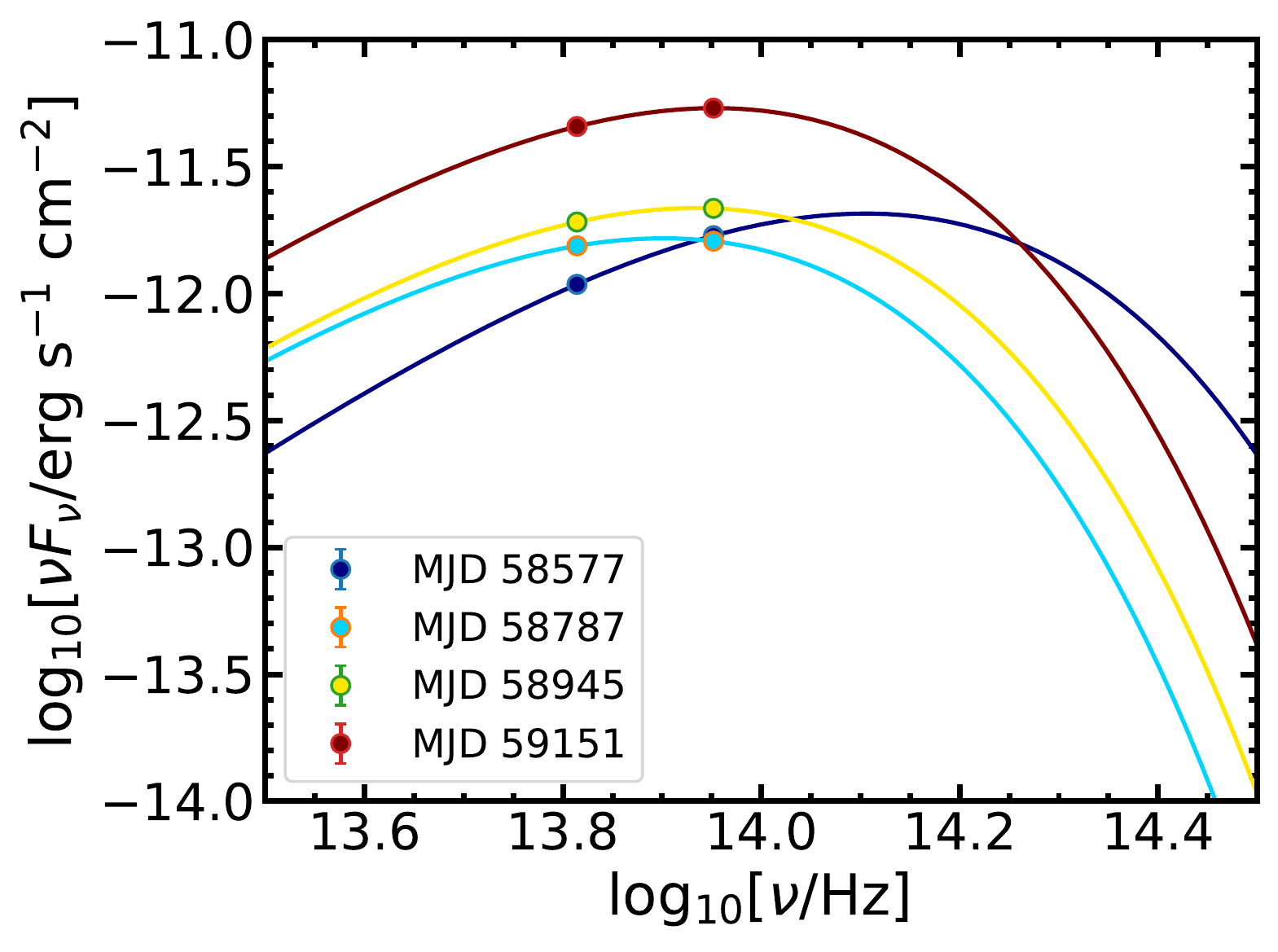}
\caption{Mid-infrared SED of AT 2019avd at different epochs of WISE observations (with the preoutburst flux removed). We fit the spectrum with the blackbodies. The best-fit parameters are shown in the Table \ref{tab:WISE_spec} with a $1\sigma$ error.}
\label{fig:WISE_spec}
\end{figure}

The mid-infrared brightening after the flare can be well explained as dust echo from TDEs \citep{Dou_Long_2016,Jiang_WISE_2016,Dou_Discovery_2017,Jiang_MIR_2017}. We apply a blackbody model to fit the mid-infrared W1 and W2 data in order to obtain the dust temperature and luminosity (see Fig. \ref{fig:WISE_spec}). The fitting results are listed in Table \ref{tab:WISE_spec}.

\begin{table}
\centering
\caption{Mid-infrared spectral fitting results with $1\sigma$ Error.}
\label{tab:WISE_spec}
\begin{tabular}{cccc}
\toprule
MJD & $T_{\rm dust} (10^3\ {\rm K})$ & $L_{\rm bb} (10^{42}\ {\rm erg\ s^{-1}})$ \\ 
\midrule
58,578 & $1.560 \pm 0.145$ & $5.70 \pm 3.59$ \\
58,788 & $0.975 \pm 0.056$ & $4.55 \pm 1.98$ \\
58,945 & $1.052 \pm 0.040$ & $5.99 \pm 1.74$ \\
59,152 & $1.099 \pm 0.026$ & $14.82 \pm 2.74$ \\
\bottomrule
\end{tabular}
\tablenotetext{}{Note. Here $T_{\rm dust}$ and $L_{\rm bb}$ are the blackbody temperature and the luminosity of the interstellar dust, respectively.}
\end{table}

\begin{figure}
\centering
 \includegraphics[scale=0.5]{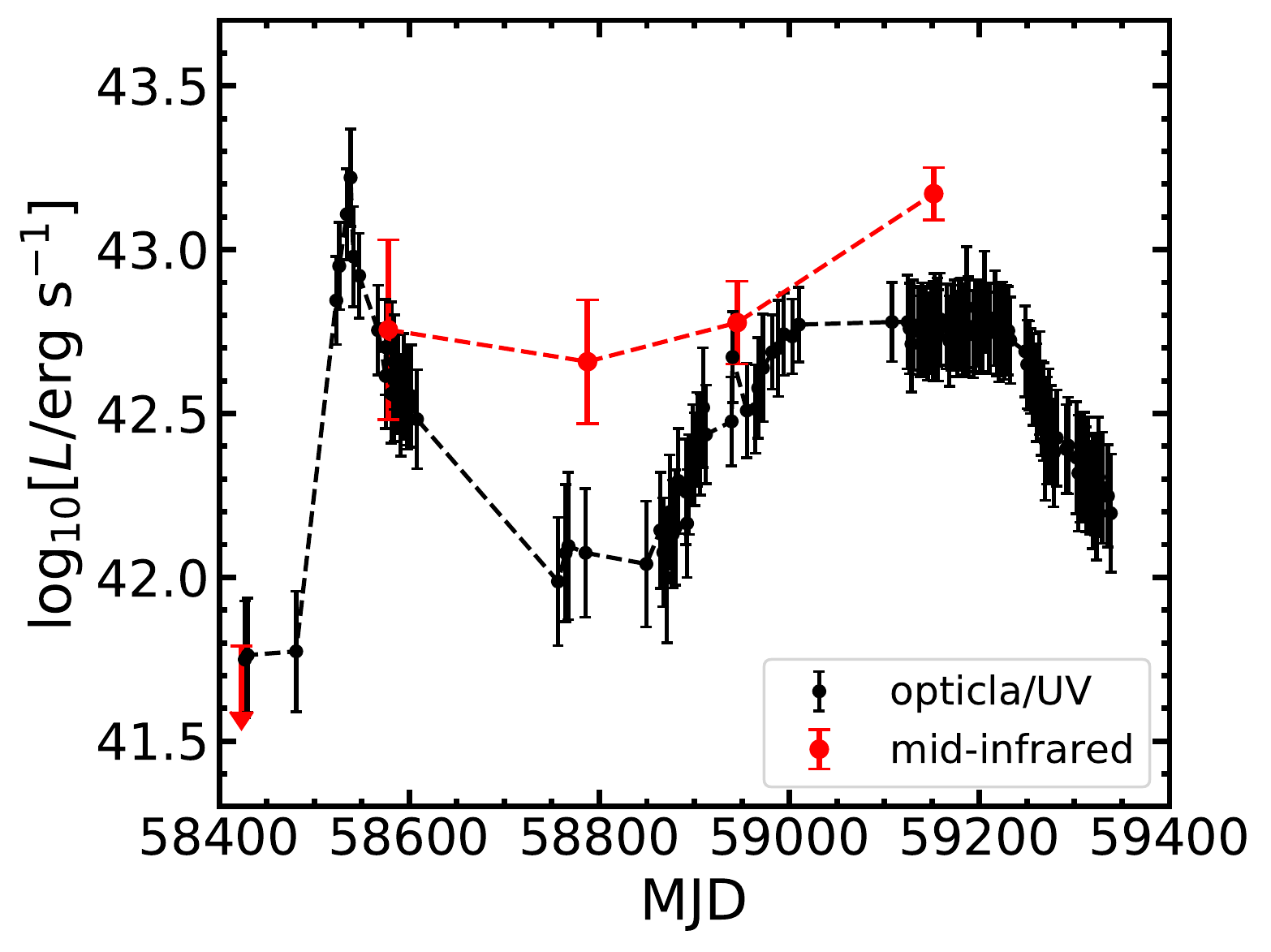}
\caption{Optical/UV and mid-infrared bolometric luminosity history of AT 2019avd. The luminosities are obtained by the spectral fitting, which is shown in Fig. \ref{fig:opt_spec} and \ref{fig:WISE_spec}. The double-peak feature in the mid-infrared light curve is associated with the optical/UV light curve. The left red down arrow represents the upper limit on MJD 58,421.}
\label{fig:opuvLC}
\end{figure}

We present the optical/UV luminosity history in Fig. \ref{fig:opuvLC} together with the mid-infrared luminosity evolution. The mid-infrared light curve also has double-peak feature similar to the optical light curve, and both of the mid-infrared peaks follow the optical ones. These features imply that the mid-infrared and the optical/UV photons are associated with the same event but come from different processes and locations.

According to the dust echo model, the mid-infrared emission comes from the interstellar dust that was heated by the UV and/or X-ray photons of a nuclear transient event \citep{Lu_Infrared_2016,Dou_Discovery_2017,Jiang_MIR_2017,Sun_A_2020}, such as a TDE. 

We can see from Fig. \ref{fig:opuvLC} that the second peak of the mid-infrared light curve seems to be brighter than the first one. It implies that the UV and X-ray photons that trigger the dust echo are more abundant near the second peak.

\subsection{X-Ray}        \label{subsec:spec_Xray}
We group the X-ray data to have at least 4 counts in each bin and mainly adopt the C-statistic for the Swift spectral fittings, which are performed using XSPEC \citep[V12.9; ][]{1996ASPC..101...17A}. We fit the X-ray spectra with the absorbed black-body model with a column density of $N_{\rm H} = 2.42 \times 10^{20}\ {\rm cm}^{-2}$ in the direction of AT 2019avd \citep[][ ,HI4PI Map]{Ben_H14PI_2016}. 

\begin{figure}
\centering
 \includegraphics[scale=0.5]{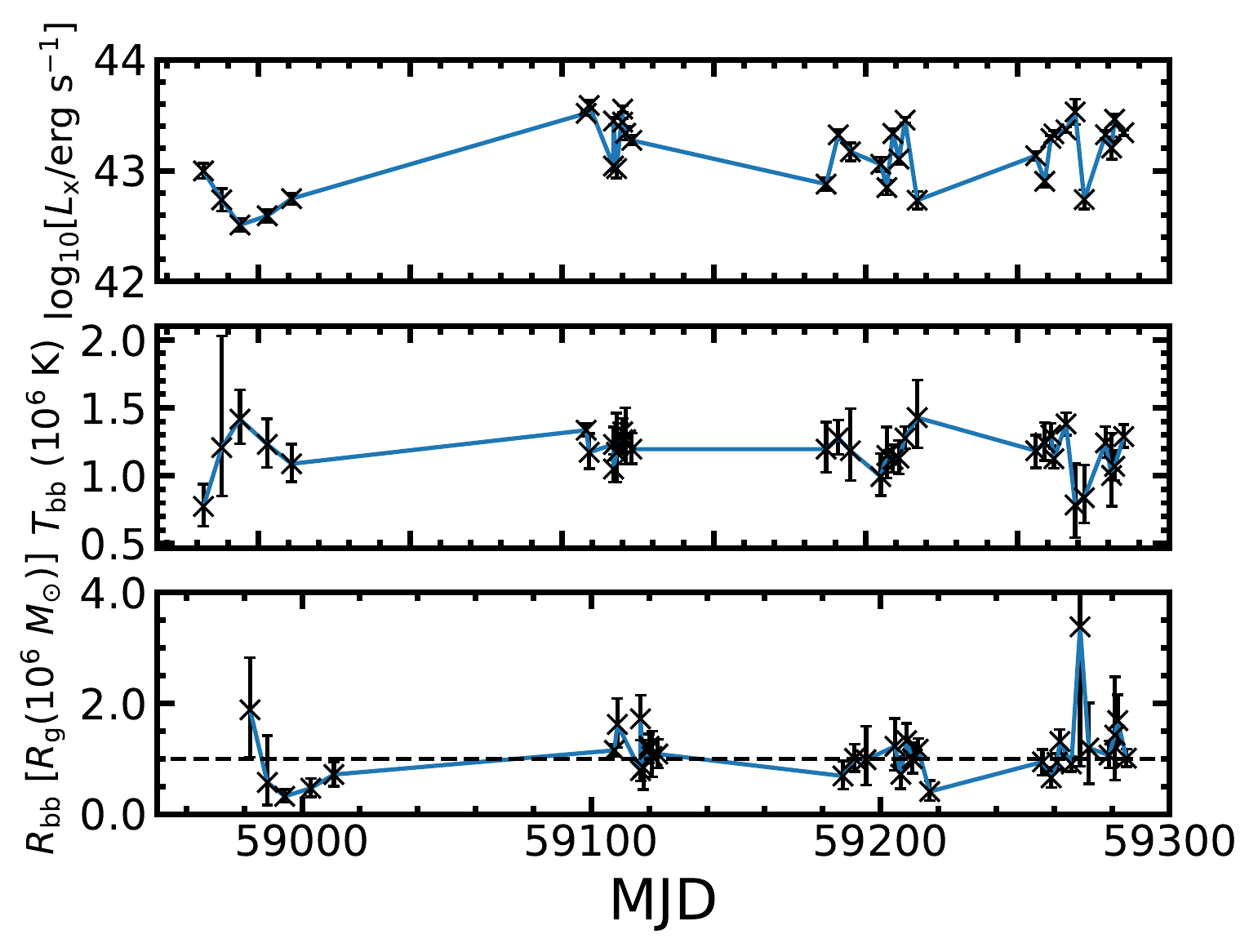}
\caption{Bolometric luminosity (top), blackbody temperature (middle), and photospheric radius (bottom) in X-ray band of AT 2019avd. The luminosity and temperature, which are listed in Table \ref{tab:xray}, are given by the blackbody spectral fit. The inferred photospheric radius is almost unchanged, with $R_{\rm bb} \sim R_{\rm g}(10^6 M_{\odot})$, which is indicated by the horizontal dashed line. It implies that the X-ray photons emit from the inner part of the accretion disk. The X-ray luminosity evolves with intensive variability that might originate from the disk precession (see the discussion in Section \ref{sec_xray_var}). Here we do not present the data after the X-ray flare (MJD $\gtrsim$ 59,368), because their fitting errors are too large.}
\label{fig:xray_reff}
\end{figure}

The fitting results are given in Table \ref{tab:xray}. We plot the X-ray luminosity $L_{\rm x} = L_{\rm bb}$, the effective temperature $T_{\rm bb}$, and the photospheric radius $R_{\rm bb} = [L_{\rm x}/(2\pi \sigma T_{\rm bb}^4)]^{1/2}$ in Fig. \ref{fig:xray_reff}. Here $\sigma$ is the Stefan--Boltzmann constant.

The X-ray luminosity evolves with intensive variability (a fluctuation factor up to $\sim 5$). We will discuss the interpretation of the X-ray variability in Section \ref{sec_xray_var}. The photospheric radius is almost unchanged with $R_{\rm bb} \sim R_{\rm g}(10^6 M_{\odot})$, where $R_{\rm g} = G M_{\rm h}/c^2$ is the gravitational radius of the SMBH. It implies that the X-ray photons come from the inner part of the accretion disk.

\section{Circularization process} \label{sec_cir}
In this and the next sections, we will demonstrate that the features of AT 2019avd can be interpreted by a two-phase evolution model of a TDE.
\subsection{Stream Crossing}			\label{subsec_SC}
In the central galaxy, the SMBH can tidally disrupt a approaching star when its pericenter $R_{\rm p}$ closes to the tidal radius  $R_{\rm T} = R_*(M_{\rm h}/M_*)^{1/3}$, i.e., the penetration factor $\beta \equiv R_{\rm T}/R_{\rm p} \sim 1$ \citep{Rees_Tidal_1988,Phinney_Manifestations_1989}. In the unit of the black hole (BH)'s Schwarzschild radius $R_{\rm S} = 2GM_{\rm h}/c^2$, the pericenter radius is
\begin{equation}
\label{eq:Rp}
R_{\rm p} \simeq 23\ \beta^{-1} M_6^{-2/3} r_*m_*^{-1/3}\ R_{\rm S}.
\end{equation}
Here $M_{\rm h} \equiv M_6 \times 10^6\ M_{\odot}$, $R_* \equiv r_* \times R_{\odot}$, and $M_* \equiv m_* \times M_{\odot}$ are the BH's mass and the star's radius and mass, respectively.

After the disruption, the stellar debris will fall back to the pericenter after the timescale \citep{Guillochon_Hydrodynamical_2013}
\begin{equation}
\label{eq:tfb}
\begin{split}
t_{\rm fb} &= 2 \pi \sqrt{a_0^3/GM_{\rm h}} \\
&\simeq 41\ M_6^{1/2} r_*^{3/2} m_*^{-1}\ {\rm day} \times \begin{cases}
\beta^{-3},&\beta \lesssim \beta_{\rm d} \\
1,&\beta > \beta_{\rm d}.
\end{cases}
\end{split}
\end{equation}
Here $a_0$ is the semi-major axis of the debris firstly returning to the pericenter, whose specific energy and eccentricity are \citep{Lodato_Stellar_2009}
\begin{equation}
\label{eq:E0}
\epsilon_0 \simeq \frac{GM_{\rm h}}{2a_0} \simeq GM_{\rm h}R_* \begin{cases}
R_{\rm p}^{-2},&\beta \lesssim \beta_{\rm d} \\
R_{\rm T}^{-2},&\beta > \beta_{\rm d}
\end{cases}
\end{equation}
and $e_0 = 1-R_{\rm p}/a_0 \simeq 1$, respectively. The critical penetration factor for full disruption is $\beta_{\rm d} \simeq 0.9\ (1.85)$ for polytropic index $\gamma = 5/3\ (4/3)$ \citep{Guillochon_Hydrodynamical_2013}.

Many studies indicate that after the disruption, the debris stream will collide with itself due to the general relativistic apsidal precession, and its orbit will circularize \citep{Rees_Tidal_1988,Hayasaki_Finite_2013,Dai_Soft_2015,Bonnerot_Disc_2016,Bonnerot_Long_2017}. The self-collision is crucial for the formation of accretion disk, however, its consequences are complicated. 

If the photons, which are induced by the self-collision, can diffuse away efficiently, the intersection can be assumed to be inelastic collision as in \cite{Dai_Soft_2015,Bonnerot_Long_2017}. The stream thus undergoes a succession of self crossings (SSC), which dissipate a significant amount of the specific energy to circularize. Eventually, the stream will settle into the circularization radius $R_{\rm c} \simeq 2R_{\rm p}$, if the stream's angular momentum is almost unchanged during the circularization process.

In the following, we will consider the SSC by analytic method, and  in Section \ref{subsec_discuss_cir} we will discuss other possible models of the circularization process. 

\subsection{Succession of Self-crossings}    \label{subsec_SSC}
In the SSC model, the stream precesses by a small angle $\phi \sim R_{\rm S} / R_{\rm p}$ upon each passage of the pericenter and thus undergoes an SSC, which dissipates an amount of the specific energy. Consequently, the apocenter of the stream's new orbit moves closer to the BH, and its eccentricity decreases. Finally, the stream is fully circularized and settles into the circularization radius $R_{\rm c}$ with the orbital energy $\epsilon_{\rm c} = GM_{\rm h} / (2R_{\rm c})$.

We adopt the differential equation (see the Eq. (9) in \cite{Chen_Light_2021})
\begin{equation}		
\label{eq:diss}
\dot \epsilon = \frac{\Delta \epsilon_0}{t_{\rm fb}} \frac{1}{e_0^2} \left(1 - \frac{\epsilon}{\epsilon_{\rm c}}\right) \left(\frac{\epsilon}{\epsilon_0}\right)^{3/2}
\end{equation}
to calculate the energy dissipation rate history. Here 
\begin{equation}		
\Delta \epsilon_0 = \frac{9}{16} \frac{\pi^2  e_0^2}{(1+e_0)^3} \left(\frac{R_{\rm S}}{R_{\rm p}}\right)^3 c^2
\label{eq:delta-eps0}
\end{equation} 
is the dissipated energy during the first crossing \citep{Dai_Soft_2015,Bonnerot_Long_2017}.

Letting $u = (\epsilon / \epsilon_0)^{-1/2}$, we can rewrite the above equation as
\begin{equation}
\dot u = \frac{1}{t_{\rm cir}} (\frac{\epsilon_0}{\epsilon_{\rm c}} u^{-2}-1).
\label{eq:dot_u}
\end{equation}
Here 
\begin{equation}
\begin{split}
t_{\rm cir} &= \frac{2 \epsilon_0 t_{\rm fb} e_0^2}{\Delta \epsilon_0} \\
&\simeq 328\ M_6^{-7/6} m_*^{-4/3} r_*^{7/2} \ {\rm day} \times \begin{cases}
\beta^{-4},&\beta \lesssim \beta_{\rm d} \\
\beta^{-3},&\beta > \beta_{\rm d}
\end{cases}
\end{split}
\label{eq:tcir}
\end{equation}
is the timescale of circularization. When $\beta > \beta_{\rm d}$, the definition of $t_{\rm cir}$ as in Eq. (\ref{eq:tcir}) is identical to that in \cite{Bonnerot_Long_2017}. We can use the following ways to comprehend $t_{\rm cir}$. In the beginning of the circularization process, $\epsilon_0/\epsilon_{\rm c} u^{-2} \sim \epsilon_0/\epsilon_{\rm c} \ll 1$, then Eq. (\ref{eq:dot_u}) reduces to $-du/dt \sim t_{\rm cir}^{-1}$. It implies that it takes $\sim t_{\rm cir}$ to circularize the stream from $u \sim 1\ (\epsilon \sim \epsilon_0)$ to $u \sim 0\ (\epsilon \sim \epsilon_{\rm c})$.

We can analytically solve the Eq. (\ref{eq:dot_u}), the solution is
\begin{equation}
\begin{split}
\frac{t-1.5t_{\rm fb}}{t_{\rm cir}} &= 1 - u + \sqrt{\frac{\epsilon_0}{\epsilon_{\rm c}}} \log \sqrt{\frac{u+(\epsilon_0/\epsilon_{\rm c})^{1/2}}{u-(\epsilon_0/\epsilon_{\rm c})^{1/2}}} \\
&= 1 - \sqrt{\frac{\epsilon_0}{\epsilon}} + \sqrt{\frac{\epsilon_0}{\epsilon_{\rm c}}} \log \sqrt{\frac{1+(\epsilon/\epsilon_{\rm c})^{1/2}}{1-(\epsilon/\epsilon_{\rm c})^{1/2}}}.
\label{eq:e_t}
\end{split}
\end{equation}
Here $t = 0$ is the disruption time, and the self-crossing occurs at $t = 1.5\ t_{\rm fb}$. Solving Eq. (\ref{eq:e_t}) by Newton's iteration, we can obtain $\epsilon(t)$. 

\begin{figure}		
\centering
 \includegraphics[scale=0.5]{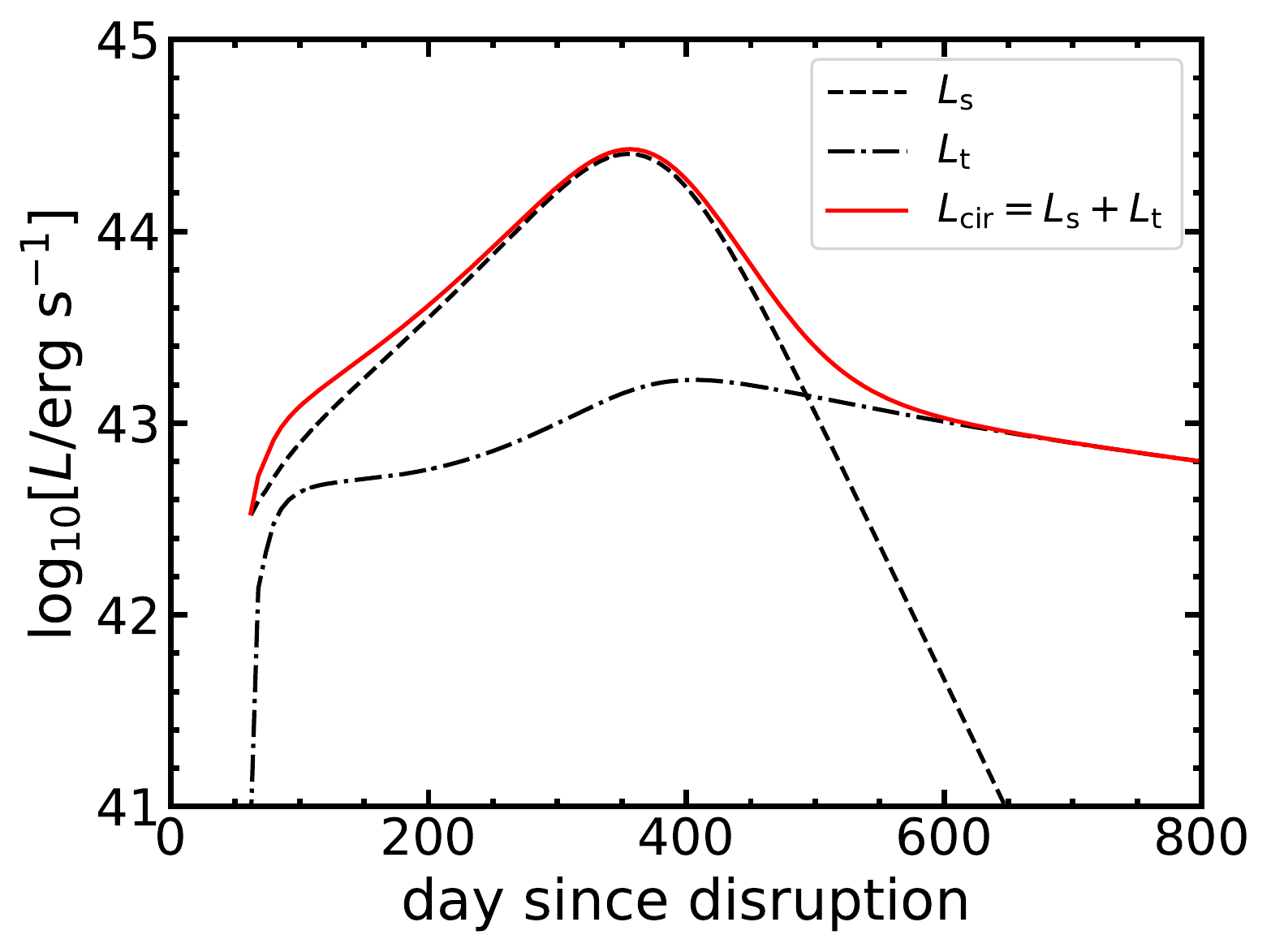}
\caption{Bolometric luminosity history during the circularization process for the disruption of a star ($m_*=r_*=1$) by a $10^6\ \rm{M_{\odot}}$ SMBH. The penetration factor is $\beta = 1$. The dashed and the dotted-dashed lines represent the luminosity of the self-collision and the tail shock, respectively. The circularization luminosity consists of both contributions.}
\label{fig:Lcir_SC}
\end{figure}

Using the differential Eq. (\ref{eq:diss}), we can write the luminosity of self-collision as 
\begin{equation}
\label{eq:Ls}
L_{\rm s}(t) \simeq M_{\rm s}(t) \dot \epsilon(t),
\end{equation}
where
\begin{equation}
\label{eq:Ms}
M_{\rm s}(t) = \int^t_{t_{\rm fb}} \dot M_{\rm fb}\ dt
\end{equation}
is the mass of the ``main stream" consisting of the total fallback mass from the beginning to now. In this paper, we adopt the following simple form of the fallback rate history:
\begin{equation}
\dot M_{\rm fb}(t) = \dot M_{\rm p} \left(\frac{t}{t_{\rm fb}}\right)^{-n}.
\end{equation}
It satisfies that $\int_{t_{\rm fb}}^\infty \dot M_{\rm fb}\ dt = \Delta M_{\rm fb}$, where $\Delta M_{\rm fb}$ is the total fallback mass. So the peak fallback rate is $\dot M_{\rm p} \simeq (n-1) \Delta M_{\rm fb}/t_{\rm fb}$. For full and partial TDEs, we let $n = 5/3$ and $9/4$, respectively \citep{Coughlin_Partial_2019,Ryu_Tidal1_2020}.

The ratio of $\Delta M_{\rm fb}$ over the total stellar mass $M_*$ depends on the stellar polytropic index $\gamma$ and the penetration factor $\beta$ \citep{Guillochon_Hydrodynamical_2013}. In this paper, we adopt the fitting formulae of the simulation results in
\cite{Guillochon_Hydrodynamical_2013} to calculate the $\beta$--$\Delta M_{\rm fb}$ relation. For a disrupted star with polytropic index $\gamma = 5/3$, the total fallback masses are $\Delta M_{\rm fb} \simeq 0.0254\ M_*$, and $0.1222\ M_*$ for $\beta = 0.6$ and $0.7$, respectively.

In reality, the fallback rate is not in a single power-law form, but it rises first and then decays in an asymptotic power law. Recent hydrodynamic simulations found that the shape of $\dot M_{\rm fb}(t)$ depends on the stellar properties, the penetration factor $\beta$, and the BH's spin \citep{Evans_The_1989,Lodato_Stellar_2009,Guillochon_Hydrodynamical_2013,Gafton_Tidal_2019,Golightly_Tidal_2019}. However, in the beginning of the self-crossing succession, the main stream already has most of the mass of the fallback debris, i.e., $M_{\rm s}(1.5t_{\rm fb}) \sim \Delta M_{\rm fb}$. Thus, the specific shape of $\dot M_{\rm fb}$ is unimportant in our calculation. In this paper, we set the power-law decay index as $n=5/3$ and $9/4$ for $\beta \gtrsim \beta_{\rm d}$ and $\beta < \beta_{\rm d}$, respectively, which correspond to the full TDE and the partial TDE regimes, respectively \citep{Lodato_Stellar_2009,Golightly_Tidal_2019}.

Besides the self-collision of the main stream, the tail of the returning stream will join the main stream during the circularization process. The luminosity of this tail shock is 
\begin{equation}
\label{eq:Lt}
L_{\rm t} \simeq \dot M_{\rm fb}(t) [\epsilon(t) - \epsilon_{\rm t}(t)].
\end{equation}
Here $\epsilon_{\rm t} (t)$ is the specific energy of the tail of the returning stream; it is given by the energy--period relation, i.e.,
\begin{equation}
\label{eq:Et}
\epsilon_{\rm t} \simeq \frac12 \left(\frac{2\pi GM_{\rm h}}{t-0.5t_{\rm fb}}\right)^{2/3},
\end{equation}
where the factor $0.5\ t_{\rm fb}$ is given by the fact that $L_{\rm t} = 0$ at the beginning of the self-crossing.

We numerically calculate the circularization luminosity $L_{\rm cir} = L_{\rm s} + L_{\rm t}$ as a function of time and plot it in Figure \ref{fig:Lcir_SC}.

By neglecting the tail shock term $L_{\rm t}$ and letting the mass of the main stream to be the total fallback mass, i.e., $M_{\rm s} \sim \Delta M_{\rm fb}$, we can estimate the peak luminosity of the circularization by
\begin{equation}
\begin{split}
L_{\rm p} &\simeq 3.1 \times 10^{44} \beta^{9/2} M_6^2 m_*^{5/2} r_*^{-9/2}\ {\rm erg/ s^{-1}} \\
&\times \begin{cases}
\frac{\Delta M_{\rm fb}}{M_*/2},&\beta \lesssim \beta_{\rm d} \\
1,&\beta > \beta_{\rm d}
\end{cases}
\end{split}
\label{eq:Lp}
\end{equation}
which is derived from Eq. (\ref{eq:E0}), (\ref{eq:diss}) and (\ref{eq:delta-eps0}) (also see the Eq. (12) in \cite{Chen_Light_2021}).

Notice that Eq. (6) of \cite{Piran_Disk_2015}, i.e., $L_{\rm p} \simeq G M_{\rm h} \dot M_{\rm p}/a_0$, also predicts the peak dissipation rate, which is different from what we obtain here. Their formula is proportional to $\beta^5$ for partial disruptions and is independent of $\beta$ for full disruptions. The latter case is different from what we give here, i.e.,  $L_{\rm p} \propto \beta^{9/2}$, although they are similar for partial TDEs. This is because they assume that the shock, occurring at the apocenter of the most bound debris, dissipates most of the gas kinetic energy. This is not always true, since the actual self-intersection radius is smaller than the apocenter radius. If the relativistic apsidal precession is large, the self-intersection radius will be closer to the pericenter radius, and more kinetic energy will be dissipated. So we consider our equation to be more general and robust.

\section{Delayed Accretion Process} \label{sec_acc}
After the circularization process, the gas will settle to the vicinity of the BH and form an accretion disk. In the accretion stage, the inflow gas and the tail of the fallback gas will supply to the accretion disk. 

Such an accretion process can be considered to be a delayed accretion. The mean accretion rate is $\dot M_{\rm acc} \simeq M_{\rm d}/t_{\rm acc}$, where $M_{\rm d}$ and $t_{\rm acc}$ are the disk mass and the accretion timescale, respectively. The change rate of the disk mass is $d M_{\rm d}/ dt = \dot M_{\rm sup}-\dot M_{\rm acc}$, where $M_{\rm sup}$ is the mass supply rate onto the disk. The solution of this equation can be written as \citep{Kumar_Mass_2008,Lin_A_2017,Chen_Tidal_2018,Mockler_Weightng_2019}
\begin{equation}
\label{eq:Macc}
\dot M_{\rm acc}(t)=\frac{1}{t_{\rm acc}} \left(\ue^{-t/ t_{\rm acc}}\int_{t_{\rm d0}}^t \ue^{t'/ t_{\rm acc}} \dot M_{\rm sup}(t')\ dt' \right),
\end{equation}
where $t_{\rm d0}$ is the starting time of the mass supply.

The supplied mass includes the main body of the bound debris, which flows to the vicinity of the SMBH and the tail of the fallback stream. The process of supply is complicated, so we simply approximate the mass supply rate as a constant, i.e., 
\begin{equation}
\label{eq:Msup}
\dot M_{\rm sup}(t) \simeq \begin{cases}
\Delta M_{\rm sup}/t_{\rm sup},&\quad t_{\rm d0} \lesssim t \lesssim t_{\rm d0} + t_{\rm sup} \\
0,&\quad {\rm otherwise}.
\end{cases}
\end{equation}
Here $\Delta M_{\rm sup}$ is the total supplied mass, and $t_{\rm sup}$ is the timescale of the mass supplying to the accretion disk still in formation. Because the circularization is prolonged and the accretion is delayed, the mass supply rate history curve would be flattened relative to the fallback rate curve. We expect that the supply timescale is longer than the fallback timescale, i.e., $t_{\rm sup} > t_{\rm fb}$.

Thus, we obtain the accretion rate
\begin{equation}
\label{eq:dotMacc}
\dot M_{\rm acc}(t) \simeq \begin{cases}
\frac{\Delta M_{\rm sup}}{t_{\rm sup}} \left(1-\ue^{-\frac{t-t_{\rm d0}}{t_{\rm acc}}} \right), \quad \quad t_{\rm d0} \lesssim t \lesssim t_{\rm d0} + t_{\rm sup} \\
\frac{\Delta M_{\rm sup}}{t_{\rm sup}} \left(1-\ue^{-\frac{t_{\rm sup}}{t_{\rm acc}}} \right) \ue^{-\frac{t-t_{\rm d0}-t_{\rm sup}}{t_{\rm acc}}}, t \gtrsim t_{\rm d0} + t_{\rm sup}.
\end{cases}
\end{equation}
The accretion luminosity is $L_{\rm acc}(t) \simeq \eta \dot M_{\rm acc}(t) c^2$, where the efficiency is taken to be $\eta = 1/12$ for a Schwarzschild BH.

\section{Model fitting to the optical/UV luminosity history} \label{sec_fitting}
In most of the TDE theories, the optical/UV radiation comes from the shocks in the circularization process, and in the accretion process, it comes from the reprocessing process and/or from the outer regions of the accretion disk \citep{Loeb_Optical_1997,Strubbe_Optical_2009,Metzger_A_2016,Dai_Unified_2018}. 

Here we fit the overall optical/UV luminosity history of AT 2019avd with our two-phase evolution model. Note that here we do not attempt to investigate the detailed mechanism/origin of the optical/UV emission during the accretion phase (e.g., whether an outflow exists and how its reprocessing produces the second peak, or how large the disk's radial extension is), and we do not put the X-ray data into the fitting because their variability will bring large uncertainties into the fit results. We will discuss the possible origin of the X-ray variability in Section \ref{sec_xray_var}.

We fit the first peak of the light curve (MJD $\lesssim 58,840$) by the SSC model and the second peak of the light curve (MJD $\gtrsim 58,840$) by the delayed accretion model. In the fitting procedure, we use the mass-radius relation of a main-sequence star \citep{Kippenhahn_Stellar_1994}
\begin{equation}
\label{eq:rstar}
r_* = \begin{cases}
m_*^{0.8},&\quad{0.08 < m_* < 1} \\
m_*^{0.6},&\quad{1 < m_* <10}.
\end{cases}
\end{equation}

\begin{figure*}
\centering
\begin{minipage}[t]{0.48\textwidth}
\centering
 \includegraphics[scale=0.37]{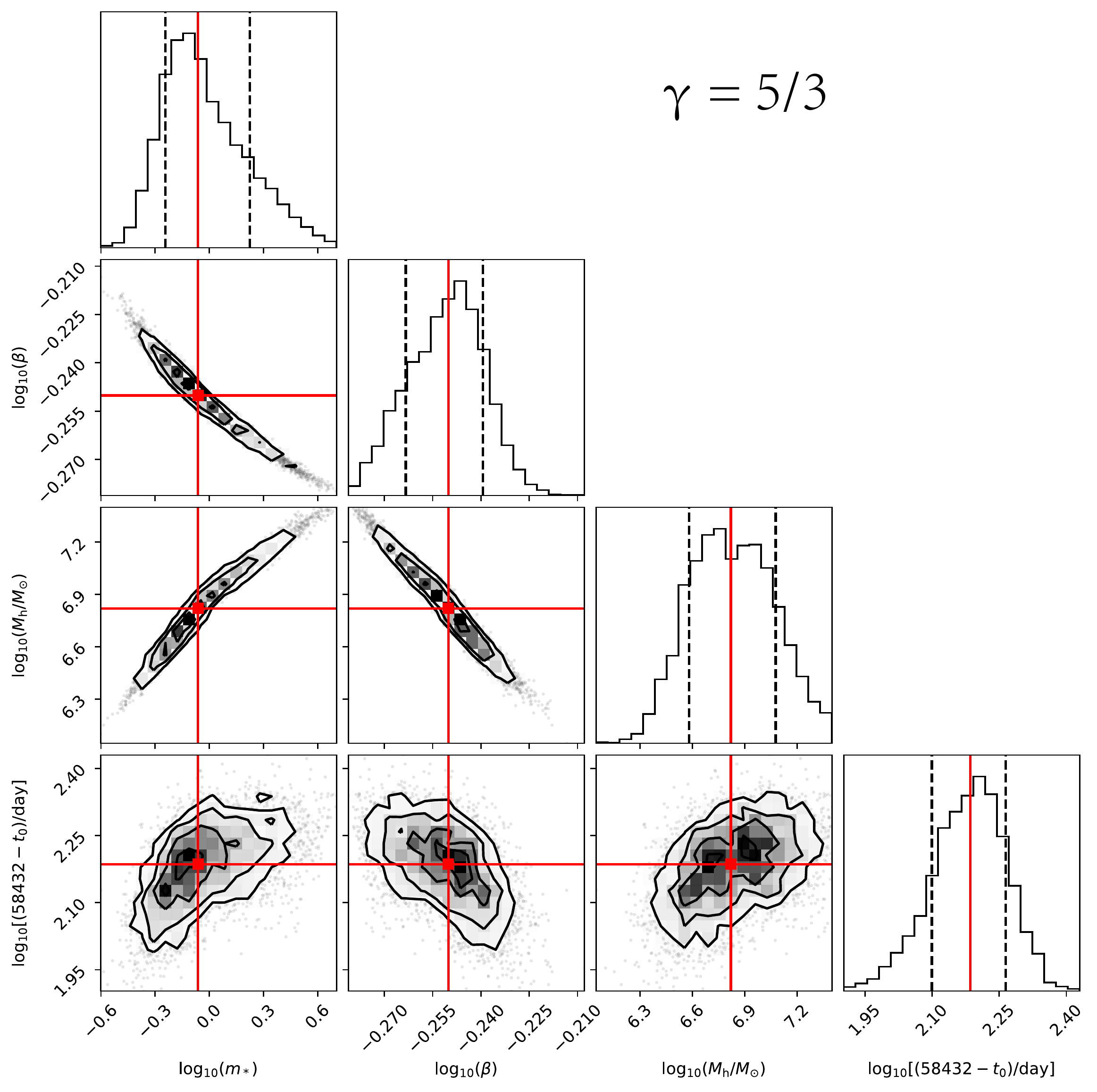}
 \end{minipage}
 \begin{minipage}[t]{0.48\textwidth}
\centering
 \includegraphics[scale=0.37]{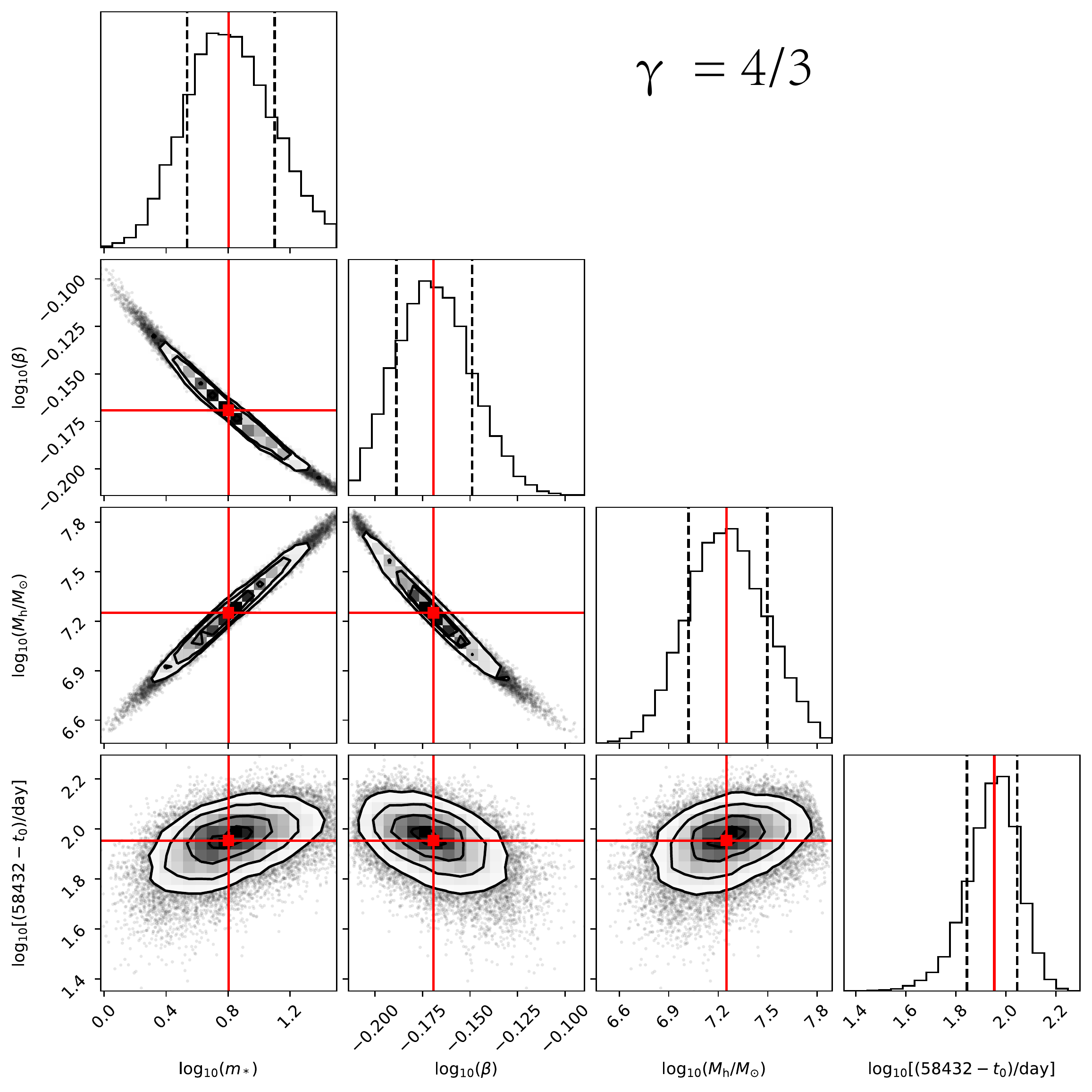}
 \end{minipage}
 \begin{minipage}[t]{0.48\textwidth}
\centering
 \includegraphics[scale=0.37]{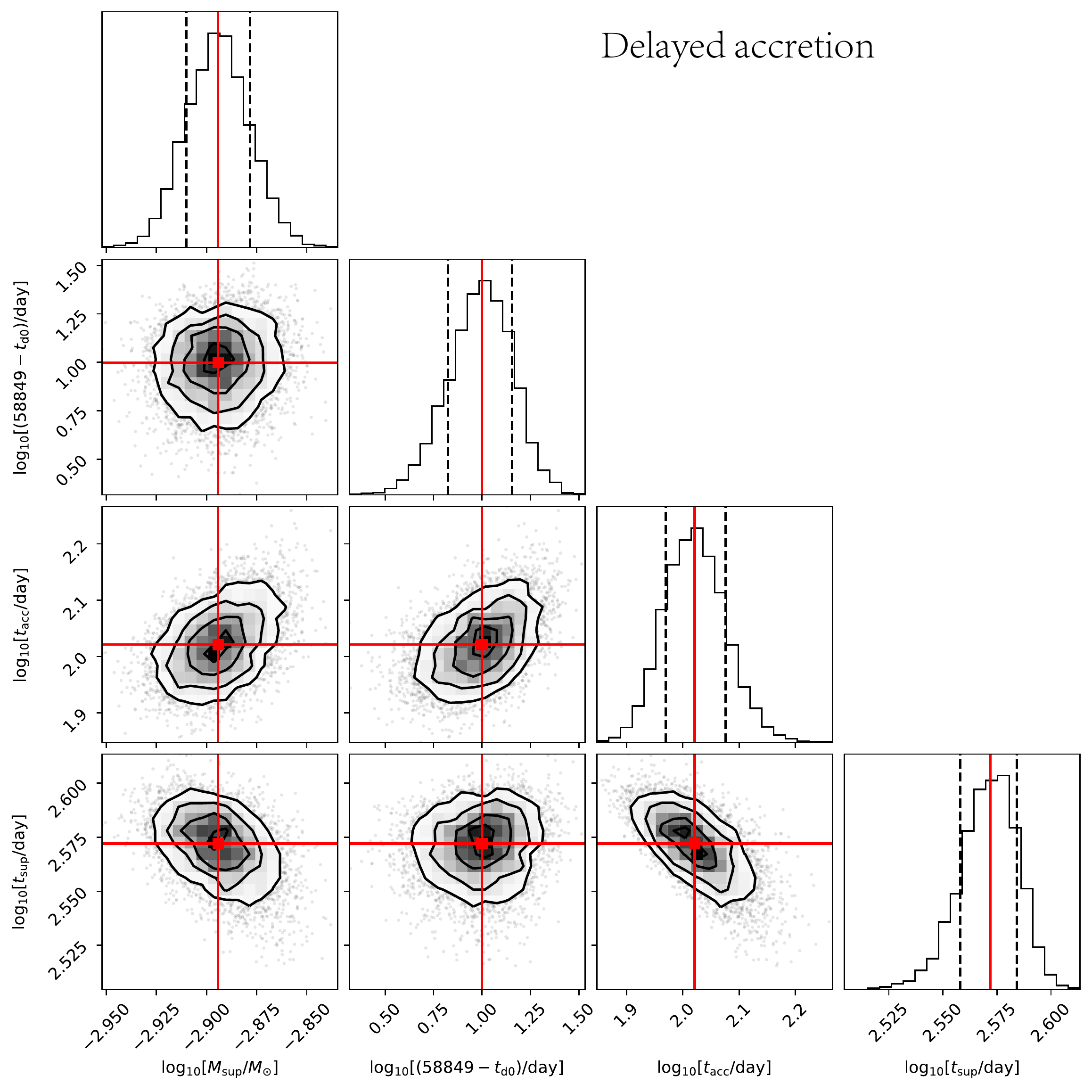}
 \end{minipage}
\caption{Posterior distribution of model parameters in the fits of AT 2019avd for the circularization phase with $\gamma = 5/3$, $\gamma = 4/3$, and the delayed accretion phase. The fitting results with a $1\sigma$ error are shown in Table \ref{tab:fitting}. Wwe plot the  corresponding light curves in Fig. \ref{fig:LC_fit}.}
\label{fig:corner}
\end{figure*}

\begin{figure}
\centering
\includegraphics[scale=0.5]{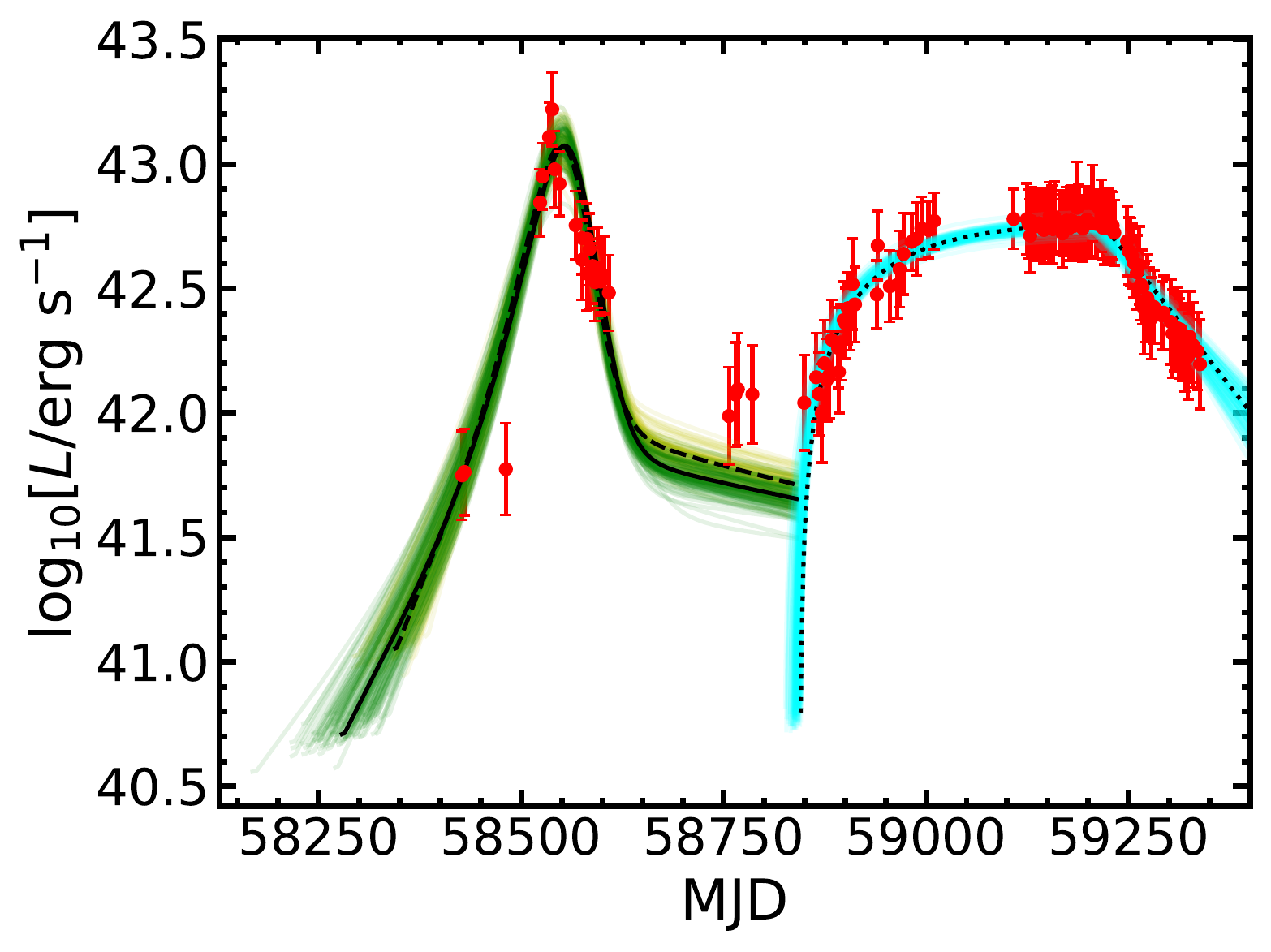}
\caption{Optical/UV bolometric luminosity history of AT 2019avd and the fitting light curves. We fit the two peaks by the SSC model (left) and the delayed accretion model (right). The best-fit light curves are represented by the black solid ($\gamma = 5/3$), dashed ($\gamma = 4/3$), and dotted lines (delayed accretion), with the distributions shown by the green, yellow, and cyan shaded areas, respectively.}
\label{fig:LC_fit}
\end{figure}

A variant of the emcee ensemble-based Markov Chain Monte Carlo (MCMC) routine \citep{Goodman_affine_MCMC_2010,Foreman-Mackey_emcee_2013} is employed. We use 40,000 steps with 30 walkers and 5000 steps to fit the circularization and accretion phases, respectively. The posterior distribution of the model parameters in the fits are shown in Fig. \ref{fig:corner}. The optical/UV luminosity history and the fitting light curves are shown in Fig. \ref{fig:LC_fit}. The fitting results are listed in Table \ref{tab:fitting}. 

\begin{table*}
\centering
\caption{Model fitting results.}
\label{tab:fitting}
\begin{tabular}{lcccccccc}
\toprule
Model & $M_*$ & $\beta$ & $M_{\rm h}$ & $t_0$ & $\Delta M_{\rm sup}$ & $t_{\rm d0}$ & $t_{\rm acc}$ & $t_{\rm sup}$ \\ 
& ($M_{\odot}$) & &  ($10^6\ M_{\odot}$) & (MJD) & ($10^{-3}\ M_{\odot}$) & (MJD) & (day) & (day) \\
\midrule
SSC ($\gamma = 5/3$) & $0.87_{-0.36}^{+0.57}$ & $0.56_{-0.02}^{+0.01}$ & $6.61_{-3.65}^{+3.91}$ & $58278_{-30}^{+27}$ & & & &  \\
SSC ($\gamma = 4/3$) & $6.35_{-3.91}^{+4.35}$ & $0.68_{-0.03}^{+0.03}$ & $17.79_{-9.44}^{+10.12}$ & $58342_{-21}^{+20}$ & & & &  \\
\midrule
Delay accretion & & & & & $1.28_{-0.05}^{+0.05}$ & $58839_{-4}^{+3}$ &  $105_{-13}^{+13}$ & $373_{-12}^{+10}$ \\
\bottomrule
\end{tabular}
\tablenotetext{}{Note. The fitting parameters on the second through ninth columns are stellar mass, penetration factor, BH mass, start time of self-crossing, total supplied mass in the accretion phase, start time of accretion, accretion timescale, and mass supply timescale, respectively.}
\end{table*}

The fit results imply that AT 2019avd is a TDE candidate that is a partial disruption of an $\sim 0.87 M_{\odot}$ star by an $\sim 6.6 \times 10^6 M_{\odot}$ SMBH with $\beta \simeq 0.56$ for $\gamma = 5/3$. This BH mass is between the two estimates derived through the H$\beta$ emission ($\sim 10^6 M_{\odot}$) and the host galaxy luminosity ($\sim 10^7 M_{\odot}$), respectively, in \cite{Frederick_A_2021}, with statistical and systematic uncertainties of 0.3--0.5 dex (also see the estimate by H$\beta$ emission in \cite{Malyali_AT2019avd_2021}, who gave $2 \times 10^6 M_{\odot}$ with uncertainty of $\sim 0.3$ dex).

The fitting result is sensitive to the stellar model. We also fit the model with a polytropic index $\gamma = 4/3$, which gives a partial disruption of an $\sim 6 M_{\odot}$ star by an $\sim 1.8 \times 10^7 M_{\odot}$ SMBH with $\beta \simeq 0.68$. The SMBH mass is slightly higher than the estimates in \cite{Frederick_A_2021}, although it is within the uncertainties. Furthermore, the derived stellar mass of the $\gamma = 4/3$ fit is larger than that of $\gamma = 5/3$ fit. We consider that such large stellar mass of $\gamma = 4/3$ is less likely because most of the tidally disrupted stars shall come from the lower end of the stellar mass function; the likelihood is reduced by $\sim 100$ times adopting the power-law index $-2.35$ of the Salpeter initial mass function \citep{Salpeter_IMF_1955}. Therefore, we prefer the $\gamma = 5/3$ fitting results.

However, the density profile of an $\sim 0.9\ M_{\odot}$ mass star, which is approximated by a solar-type star with $\gamma = 4/3$, is inconsistent with a priori assumption of a $\gamma = 5/3$ polytrope. The $\gamma = 4/3$ and $5/3$ polytropes have been used to describe the radiative core and the convective envelope of stars, respectively \citep{Rappaport_A_1983}. The polytropic index of a solar-type star with a radiative core is likely to be $\gamma = 4/3$. If the star is fully disrupted, $\gamma = 4/3$ would be a good approximation. However, if only the surface of the star is stripped away (partial TDE), it would be ideal to use a more accurate stellar structure, e.g., a composite polytrope model  \citep{Horedt_Polytropes_2004}, or an accurate density profile generated by a stellar evolution code, the latter of which has been used to study the fallback rate of TDEs in \cite{Golightly_On_2019}. 

Unfortunately, no empirical formula for the relation between $\Delta M_{\rm fb}$ and $\beta$ has emerged from numerical simulations adopting a composite stellar structure. In this paper, our motivation is to present that the double-peak light curve of AT 2019avd can be interpreted by the two-phase TDE model, so we consider this simple approach acceptable.

The penetration factor $\beta$ determines the pericenter radius of the star. The self-collision is stronger for larger $\beta$, causing faster circularization and higher luminosity. For $\beta < \beta_{\rm d}$, the total fallback mass is sensitive to $\beta$. Our fitting results of SSC (Table \ref{tab:fitting}) indicate that AT 2019avd is a partial disruption with $\beta \simeq 0.56$ whose total fallback mass is $\Delta M_{\rm fb} \simeq 0.02\ M_{\odot}$.

The total fallback mass is larger than the total supplied mass ($\Delta M_{\rm sup} / \Delta M_{\rm fb} \sim 0.1$). It implies that the SMBH can only accrete one-tenth of the debris. Most of the debris might become unbound during the circularization process as outflow and/or during the accretion process as disk wind. This unbound debris can reprocess some portion of the X-ray photons into the optical/UV emission.

For the delayed accretion model, we have four free parameters: total supplied mass $\Delta M_{\rm sup}$, supply timescale $t_{\rm sup}$, accretion timescale $t_{\rm acc}$ and the start time of accretion $t_{\rm d0}$. The accretion and the supply timescales determine the slope and the rise time of the luminosity history, respectively. The total supplied mass determines the total radiative energy by accretion.

The supply timescale $t_{\rm sup} \simeq 373$ days, and the accretion phase start time $t_{\rm d0} \simeq 58,839$ MJD imply that the process of the mass supply to the disk is long, and the time of accretion is delayed. After the disruption, the debris should experience a long-term circularization process if the general relativity effect is weak. For the partial disruption of a star by a modest SMBH, the general relativity effect might not be strong for the rapid formation of the accretion disk. Therefore, most of the debris will not necessarily flow into the last stable circular orbit of the SMBH at the early time, and the accretion process is delayed. The supplied mass includes the bound debris, which flows to the vicinity of the SMBH and the tail of the fallback stream. However, the details about how the mass is supplied to the accretion disk is an open question, and it needs future simulation to elucidate.

The fit of delayed accretion gives a accretion rate of $\sim \Delta M_{\rm sup}/t_{\rm acc} \sim 0.01\ \dot M_{\rm Edd}$, which is far smaller than the Eddington rate $\dot M_{\rm Edd} = L_{\rm Edd} / (\eta c^2)$: the typical accretion rate for full TDEs. This result is based on the assumption of radiatively efficient accretion, in which the efficiency of converting accretion power to luminosity is $\eta = 1/12$. Previously, radiatively inefficient accretion flows (RIAFs) had been used to explain the low-luminosity accretion disk in galactic nuclei \citep{Mahadevan_Scaling_1997}, however, we consider that our case here would not be the RIAF because the accretion rate for the RIAF needs to be $<10^{-3}\ M_{\rm Edd}$ with a viscosity parameter $\alpha \sim$ 0.1--0.3 \citep[see the Eq. (52) in][]{Mahadevan_Scaling_1997}.

Furthermore, the optical depth of the disk is $\tau \sim \kappa_{\rm es} \Sigma \sim \kappa_{\rm es} \Delta M_{\rm sup} / R_{\rm c}^2 \sim 10^3 \gg 1$, where $\kappa_{\rm es} \simeq 0.34\ \rm{cm^2\ g^{-1}}$ is the Thompson electron scattering opacity for the solar composition, and the size of the newly formed disk is $\sim R_{\rm c}$ if  the angular momentum of debris is almost unchanged during the early time. Therefore, we consider that the disk of our case should be optically thick, and it is not an RIAF.

The accretion timescale given by the fit is $t_{\rm acc} \simeq 105$ days. In theory, the accretion process is driven by the viscous shear; thus, the accretion timescale should be equal to the viscous timescale at $R_{\rm c}$, i.e., $t_{\nu}(R_{\rm c}) \simeq t_{\rm acc}$. The viscous timescale is given by $t_{\nu}(R) = R^2/\nu_{\rm vis}$. Using the $\alpha$ viscosity prescription $\nu_{\rm vis} \simeq \alpha \Omega_{\rm k}(R) H^2$ \citep{Shakura_viscosity_1973}, one can obtain
\begin{equation}
\begin{split}
t_{\rm acc} &\sim t_{\nu}(R_{\rm c}) \simeq [\alpha \Omega_{\rm k}(R_{\rm c})]^{-1}(H/R)^{-2} \\
&\simeq 74\ \left(\frac{\alpha}{0.1} \right)^{-1} \left(\frac{H/R}{0.1} \right)^{-2} \beta^{-3/2} m_*^{-1/2} r_*^{3/2}\ {\rm day},
\end{split}
\label{eq:tv}
\end{equation}
where $\Omega_{\rm k}(R_{\rm c})$ is the angular velocity at $R_{\rm c}$. 

In order to interpret the accretion timescale $t_{\rm acc} \simeq 105$ days, the scale-height ratio should be $H/R > 0.01$. The optimistic values are $\alpha = 0.3$ and $H/R \simeq 0.1$ for the fitting results. This scale-height ratio is larger than the standard geometrically thin disk that usually has $H/R \simeq 0.01$ \citep{Kato_Black_1998}. In other words, in order to explain the rapid accretion with a short accretion timescale $t_{\rm acc} \sim 105$ days, a standard geometrically thin disk that has been used in the theoretical models to predict the features of delayed accretion in TDEs \citep{Chen_Light_2021,Hayasaki_On_2021}, would not be valid here. We do not have a physically robust reason for this large scale-height ratio $H/R \simeq 0.1$. We suspect that some effects might heat up the accretion disk during the disk formation; thus, the heating raises the disk height and accelerates the accretion process. We will discuss this issue in Section \ref{subsec_discuss_accretion}.

\section{Interpretation of the X-Ray variability} \label{sec_xray_var}
Large-amplitude variabilities in accreting BH-related X-ray transients, such as seen in AT 2019avd (Section \ref{subsec:spec_Xray}) might originate from a precessing relativistic jet \citep{Lei_Frame_2013,Wang_QP_2014}
, a precessing accretion disk \citep{Stone_Observing_2012,Shen_Evolution_2014}, or the disk instability \citep{Lightman_Black_1974,Janiuk_On_2011}. The lack of hard X-ray photons detected in AT 2019avd makes it unlikely that variability is induced by jet precession, a scenario similar to the jet TDE Swift J1644+57 \citep{Burrows_Relativistic_2011}. The thermal or viscous instability in an accretion disk can induce a limit-cycle behavior in the light curve \citep{Janiuk_On_2004,Merloni_On_2006,Grzedzielski_Modified_2017}. However, such instability-induced variability often has a duty cycle $\ll 1$ that is not observed in AT 2019avd.

We consider that the X-ray variability of AT 2019avd is caused by the disk precession. It is very likely that the accretion disk is misaligned with the equatorial plane of the spinning BH, and the accretion disk will be subject to the Lense--Thirring effect \citep{Lense_LT_1918}. 

The Lense--Thirring torque has a strong radial dependence, i.e., $\propto R^{-3}$; thus, frame dragging acts most rapidly on the inner regions of the disk. The outcome of the torque depends on the structure of the accretion disk. If the disk is thin ($H/R \ll 1$), in the inner region, the warp of the disk propagates in a diffusive way and would lead to rapid alignment by the Bardeen--Petterson effect \citep{Bardeen_BPeffect_1975}. However, if the disk is thick, which is expected in TDEs \citep{Strubbe_Optical_2009}, the disk will precess as a solid-body rotator as shown in the simulations \citep{Nelson_Hydrodynamic_2000,Fragile_Hydrodynamic_2005,Fragile_Global_2007,Dexter_Observational_2011}.

The disk surface density is usually proportional to some power law of the radius, i.e., $\Sigma(R) \propto R^{-\zeta}$; thus, the period of the rigid disk precession within the outer radius $R_{\rm out}$ is given by \citep{Stone_Observing_2012,Shen_Evolution_2014}
\begin{equation}
\begin{split}
P_{\rm d} &= \frac{\pi c^3}{a G^2 M_{\rm h}^2} 
\frac{\int_{R_{\rm in}}^{R_{\rm out}} R'^{3/2} \Sigma(R')\ dR'}{\int_{R_{\rm in}}^{R_{\rm out}} R'^{-3/2} \Sigma(R')\ dR'} \\
&= P_{\rm LT} f(r_{\rm out}, \zeta).
\end{split}
\label{eq:P_d}
\end{equation}
Here
\begin{equation}
P_{\rm LT} = \frac{\pi c^3}{a G^2 M_{\rm h}^2} R_{\rm in}^3
\end{equation}
is the Lense--Thirring precession period for a gas particle located at the innermost circular orbit $R_{\rm in}$ and
\begin{equation}
\begin{split}
f(r_{\rm out}, \zeta) &= \frac{\int_1^{r_{\rm out}} r'^{3/2} \Sigma(r')\ dr'}{\int_1^{r_{\rm out}} r'^{-3/2} \Sigma(r')\ dr'} \\
&\simeq \begin{cases}
\frac{2\zeta+1}{2\zeta-5} r_{\rm out}^3,&\zeta \lesssim -1/2 \\
\frac{2\zeta+1}{2\zeta-5} r_{\rm out}^{5/2-\zeta},&-1/2 < \zeta < 5/2 \\
\frac{2\zeta+1}{2\zeta-5},&\zeta \gtrsim 5/2 ,
\end{cases}
\end{split}
\label{eq:fout}
\end{equation}
is a function of the ratio $r_{\rm out} \equiv R_{\rm out}/R_{\rm in}$ and the power-law index of the surface density $\zeta$. Here the normalized variable $r \equiv R / R_{\rm in} \gtrsim 1$. The second equation is obtained by letting $r_{\rm out} \gg 1$. The above equation implies that if $\zeta \lesssim 5/2$, $f \propto r_{\rm out}^3 \ {\rm or}\ r_{\rm out}^{5/2-\zeta}$, the outer regions of the disk contribute most of the Lense--Thirring torque; otherwise, if $\zeta \gg 5/2$ or $f \simeq 1$, the precession is dominated by the inner regions of the disk.

\cite{Shen_Evolution_2014} studied the disk evolution of TDEs. They found that the power law is $-1/2 \lesssim \zeta \lesssim 1/2$ for the advective index regime and $-3/2 \lesssim \zeta \lesssim -1/2$ for the radiative regime of the disk. The disk in either of the two regimes can always precess as a solid body. Therefore, we consider the range of $\zeta$ as $-3/2 \lesssim \zeta \lesssim 1/2$.

The inner radius of the disk depends on the spin of the SMBH and is approximated by the innermost stable circular orbit \citep[a negative/positive sign below corresponds to the prograde/retrograde orbit; ][]{Bardeen_Rotating_1972}
\begin{equation}
\begin{split}
R_{\rm in} &\simeq \left(3+ Z_2 \pm \sqrt{(3-Z_1)(3+Z_1+2Z_2)} \right) R_{\rm g} \\
&\simeq g(a) R_{\rm g},
\end{split}
\label{eq:Rin}
\end{equation}
where $Z_1 = 1+(1-a^2)^{1/3} ((1+a)^{1/3}+(1-a)^{1/3})$ and $Z_2 = (3a^2+Z_1^2)^{1/2}$. For the prograde and retrograde orbits with maximum spin $a = 1$, we have $g(a) = 1$ and $9$, respectively.

\begin{figure}
\centering
 \includegraphics[scale=0.5]{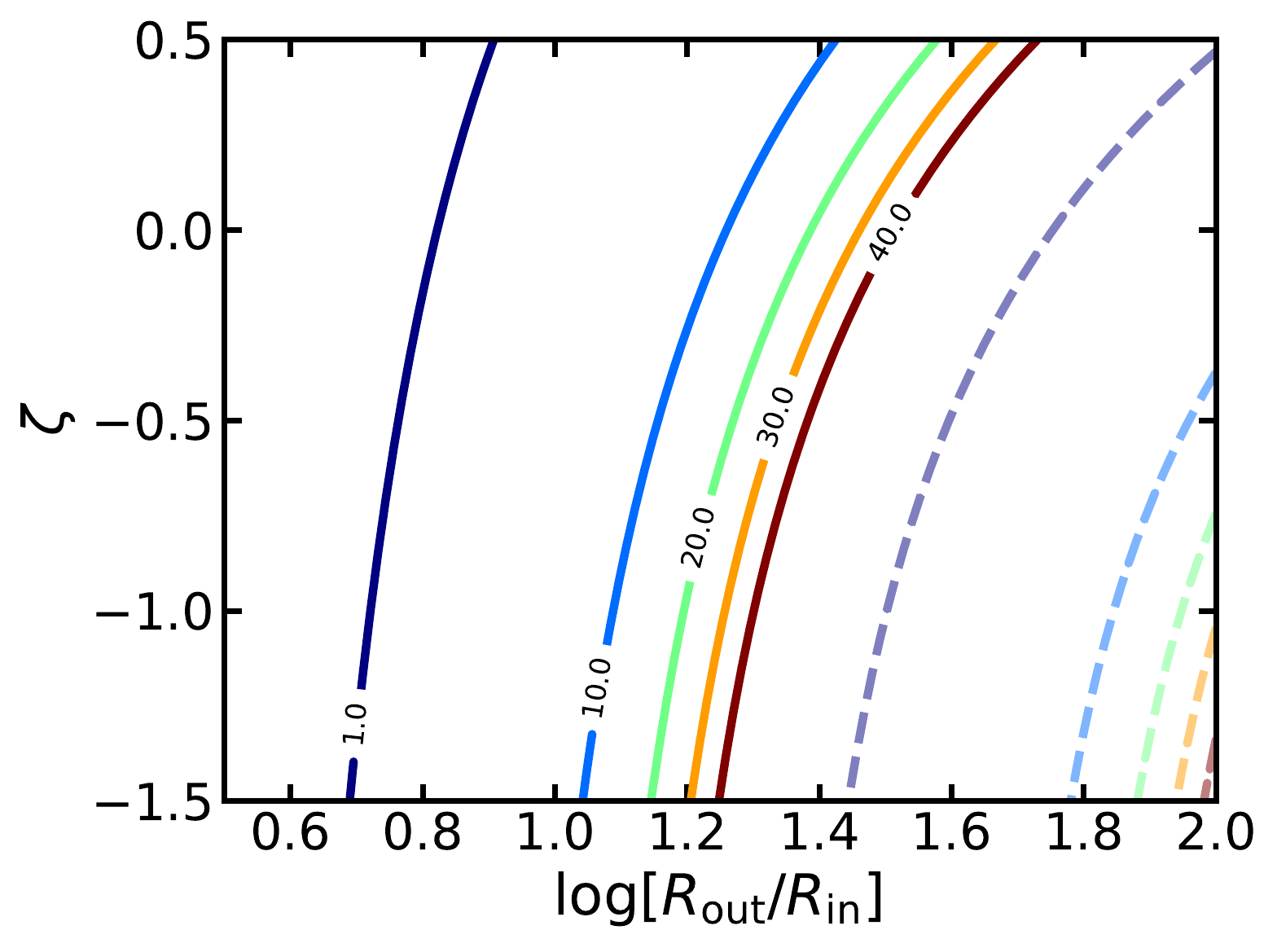}
\caption{Period of the disk precession as a solid body with respect to the power law $\zeta$ of the disk surface density and the ratio $R_{\rm out}/R_{\rm in}$ of the precessing disk. We use different colored lines to represent the period, with the values $P_{\rm d}$ in units of days shown along the lines. The solid and the dashed lines represent the BH's spins $a = 0.5$ and $a= 1$, respectively. The period is given by Eq. (\ref{eq:P_d_nor}) with a BH's mass of $10^6 M_{\odot}$.}
\label{fig:LT_period}
\end{figure}

Substituting Eq. (\ref{eq:fout}) and (\ref{eq:Rin}) into Eq. (\ref{eq:P_d}), one can rewrite the period of the disk precession as
\begin{equation}
P_{\rm d} \simeq 1.8\ M_6 \frac{g(a)^3}{a} \frac{f(r_{\rm out}, \zeta)}{10^4}\ {\rm day}.
\label{eq:P_d_nor}
\end{equation}
Here $M_{\rm h} \equiv M_6 \times 10^6\ M_{\odot}$ is the BH's mass. We plot the period of the disk precession $P_{\rm d}$ in Fig. \ref{fig:LT_period} with respect to the power-law index $\zeta$ and the ratio $r_{\rm out}$.

The period of the disk precession in AT 2019avd is $\sim$ 10--25 days. We cannot obtain the precise period due to long gaps in the X-ray data for frequency analysis. This estimate of the period is supported by the simulation results of \cite{Hayasaki_spin_2016}. We can infer from Fig. \ref{fig:LT_period} that the outer radius of the precessing disk is $R_{\rm out} \sim {\rm few} \times 10\ R_{\rm in}$, if the BH mass is $10^6\ M_{\odot}$ with moderate spin $a = 0.5$. This ratio of the radius $R_{\rm out}/R_{\rm in}$ is consistent with the size of the accretion disk expected in TDEs \citep{Strubbe_Optical_2009,Shen_Evolution_2014}. 

\begin{figure}
\centering
 \includegraphics[scale=0.5]{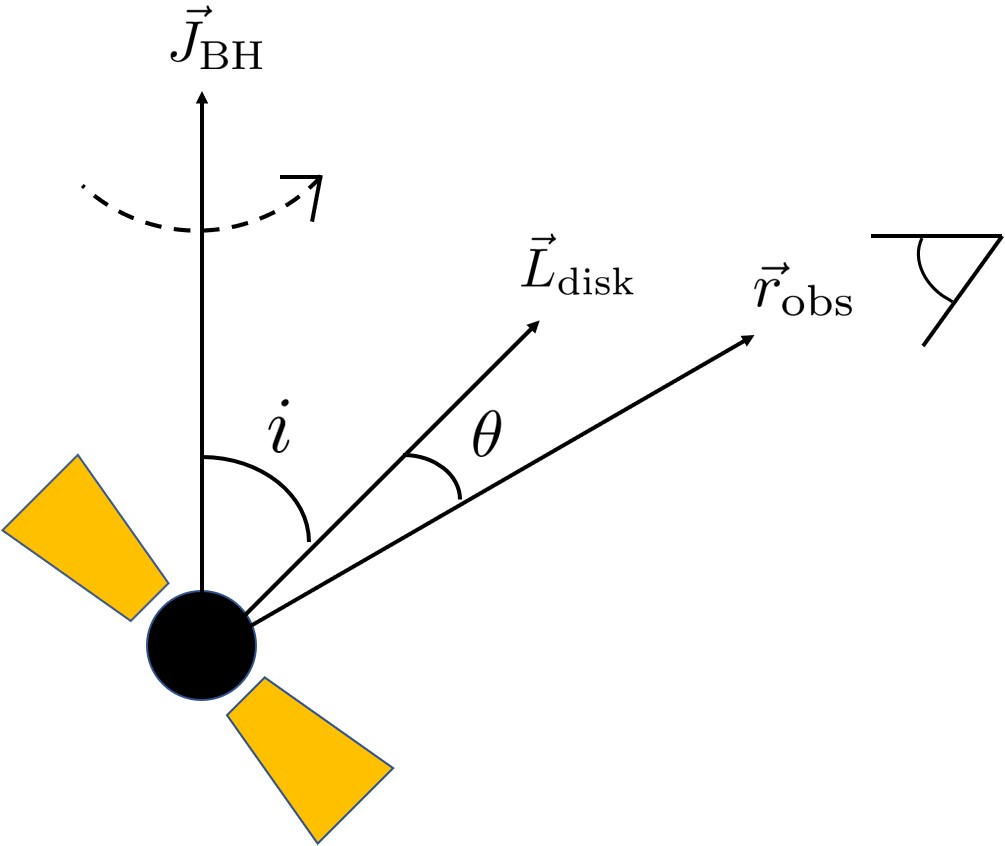}
\caption{Sketch of the disk precession. The angle $i$ between the SMBH spin vector $J_{\rm BH}$ and the disk angular momentum vector $L_{\rm disk}$ is a constant. The angle $\theta$ between the line of sight $r_{\rm obs}$ and the disk orientation changes as the disk precesses along the SMBH spin vector.}
\label{fig:LT}
\end{figure}

The amplitude of the X-ray variation depends on the line of sight, the SMBH spin, and the disk orientation, as illustrated in Fig. \ref{fig:LT}. As the disk precesses, the observed flux evolves with the angle $\theta$ ($\propto \cos{\theta}$) between the line of sight and the disk orientation. Here we only consider the small inclination, i.e., $2i + \theta_0 < \pi/2$, where $\theta_0$ is the minimum $\theta$. Otherwise, the X-ray emission will disappear as the disk precesses to the edge-on position. When $\theta = \theta_0$ and $\theta = 2i + \theta_0$, the observed flux reaches its maximum and minimum, respectively. The amplitude of the X-ray variation can be expressed by the ratio of the maximum flux and the minimum flux, i.e.,

\begin{equation}
A \simeq \frac{\cos \theta_0}{\cos (2i+\theta_0)}.
\label{eq:amplitude}
\end{equation}
Because the X-ray flux of AT 2019avd is bright ($\sim 10^{43}\ {\rm erg\ s^{-1}}$), we can assume the disk is almost face-on, i.e., $\theta_0 \sim 0$, and obtain $A \sim 1 / \cos (2i)$. For AT 2019avd, $A \sim 5$, the angle between the SMBH spin vector and the disk angular momentum vector is $i \sim 39\degree$.

\section{Conclusions and Discussion} \label{sec_conclus}
The TDE candidate AT 2019avd has double peaks in its optical light curve, and the Bowen fluorescence lines and the X-rays emerge near its second peak. These features raise a question: whether its two peaks belong to different processes. We fit the SEDs of AT 2019avd in the optical/UV, mid-infrared, and X-ray bands, then plot their corresponding luminosity histories. We find that the mid-infrared luminosity has double peaks similar to the optical/UV light curve, and its second peak seems to be brighter than the first one.

The presence of Bowen fluorescence lines, the detection of the X-ray, and the renewed and enhanced second mid-infrared rise all the three features happen near the second optical peak. This implies a delayed formation of the accretion disk around that time. We consider that the first and the second peaks come from the circularization and the delayed accretion process, respectively.

We use the circularization process plus the delayed accretion to interpret the double-peak feature in the light curve of AT 2019avd. The fitting results are consistent with the partial disruption of a $0.87 M_{\odot}$ star by a $7 \times 10^6 M_{\odot}$ SMBH with $\beta \simeq 0.56$. Therefore, AT 2019avd can be interpreted as a partial TDE.

We analyze the X-ray properties and find that the X-ray photons come from the inner part of the accretion disk, and the luminosity evolves with intensive variability (fluctuation factors up to $\sim 5$). This variability can be explained by the rigid disk precession, whose period is $\sim 10 - 25$ days (see the discussion in Section \ref{sec_xray_var}). To our knowledge, it is the first ever non-jetted TDE candidate that has shown putative signs of disk rigid-body precession.

The disk precession might diminish due to the alignment process. \cite{Franchini_LT_2016} considered that the timescale of the alignment depends on the viscosity parameter and the BH's mass and spin. If the X-ray variability originates from the disk precession, we expect that the X-ray variability will diminish within $10^3$ days, with the BH's mass $\sim 10^6 M_{\odot}$ and moderate spin $a \lesssim 0.5$.

\subsection{Circularization in Partial TDEs} \label{subsec_discuss_cir}
If the general relativity effect is weak, the debris cannot rapidly form an accretion disk; instead, the debris will experience a long-term circularization process until it has have lost enough energy and angular momentum for accretion \citep{Hayasaki_Finite_2013,Bonnerot_Disc_2016,Bonnerot_Long_2017}. In this case, i.e., partial TDE, the pericenter of the star is far from the SMBH, which weakens the self-collision and causes delayed formation of the accretion disk. 

An elliptical disk might form before the stream's orbit is completely circularized \citep{Svirski_Elliptical_2017,Lynch_Dynamical_2021}, so the accretion process would start in advance. However, it is unclear whether an elliptical disk can persist while the debris in different orbits moves with relativistic apsidal precession and if the thermal pressure and the viscous shear can efficiently redistribute the angular momentum of the debris for accretion during the early circularization process. Despite this unresolved issue, for a partial TDE, we expect that the disk formation is delayed due to the inefficient circularization process \citep{Chen_Light_2021}. Notice that \cite{Hayasaki_On_2021} also considered the delayed accretion model but in full TDEs.


The SSC circularization model we use to fit the data is based on the assumption that the photons induced by the self-collision can diffuse away efficiently. It may be applicable for a partial TDE; however, if the photon cannot diffuse efficiently, the postshock gas will be pushed away by the radiation pressure. In this case, the consequences are more complicated. In the simulation of \cite{Bonnerot_Simulating_2020} and \cite{Lu_Self_2020}, they considered that if the self-collision is intensive due to the strong general relativity effect, some of the postshock gas will be unbound, which is the so-called collision-induced outflow, and some of the bound gas will flow into the vicinity of the BH and produce the second shock. The second shock, which occurs near the BH, is crucial for the formation of the accretion disk and also the main source of radiation. As the mass inflow rate follows the mass fallback rate, the luminosity should follow the fallback rate to decay as a power law in the late time. 

%

\begin{figure}
\centering
 \includegraphics[scale=0.5]{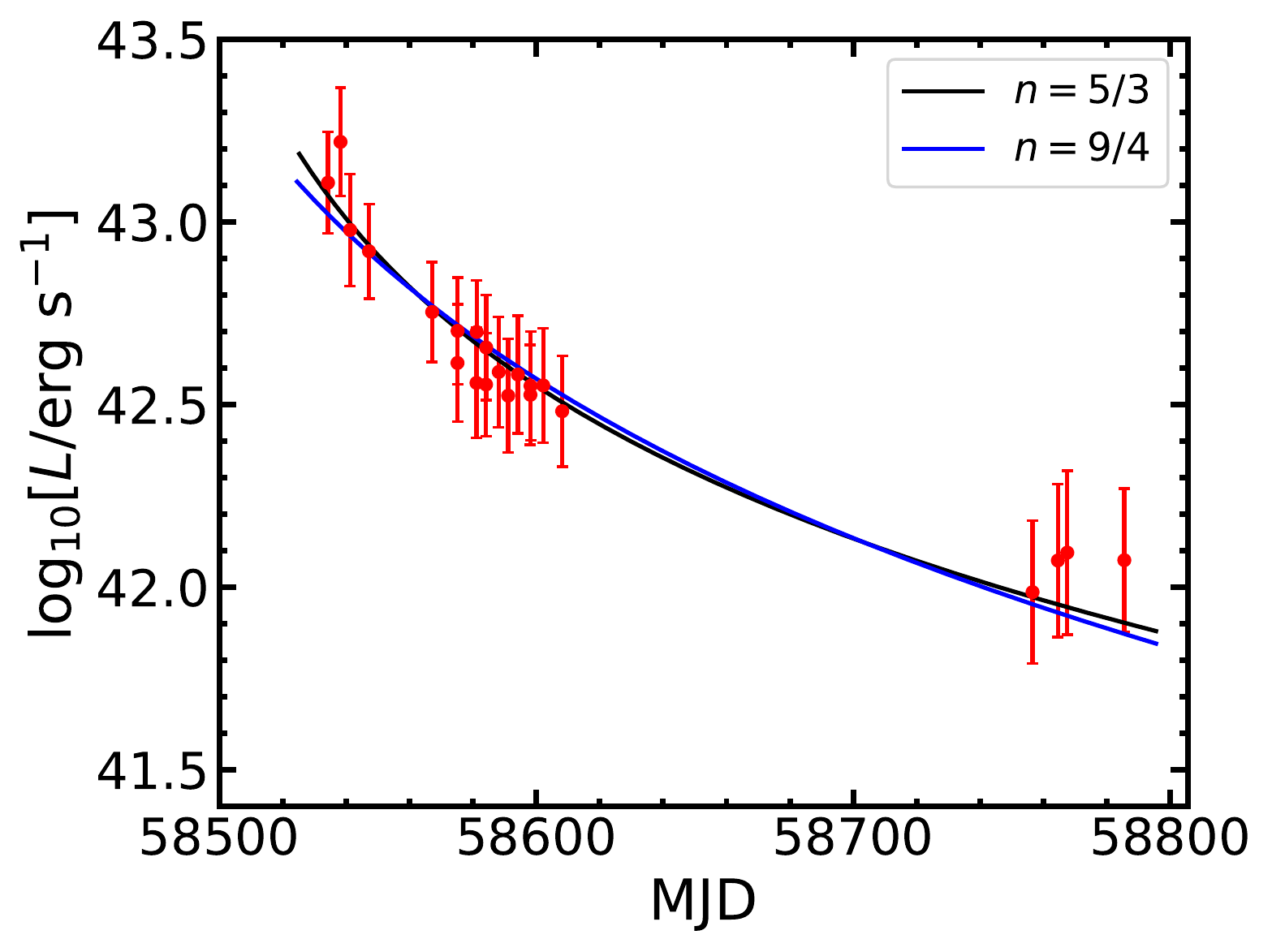}
\caption{Power-law fit of the decay of the first peak in the optical/UV bolometric luminosity. The decay can fit well by the power-law decays.}
\label{fig:PL_fit}
\end{figure}

In Fig. \ref{fig:LC_fit}, the SSC model can well fit the light curve, except for the late time of the circularization phase. The decay after the first peak looks like a power-law decay, which is similar to the scenario discussed above. Thus, we use a power-law decay $(t/t_{\rm p})^{-n}$ to fit the decay part of the light curve during the circularization phase, as shown in Fig. \ref{fig:PL_fit}. For full and partial TDEs with $n = 5/3$ and $9/4$, the decay timescales are $t_{\rm p} \simeq 53$ and $102$ days, respectively. The decay timescales are consistent with the typical TDE fallback timescale (see Eq. (\ref{eq:tfb})).

The circularization process still needs to be confirmed by the more robust radiative hydrodynamic simulations, especially for the partial TDE. And we can use the observations to check the theory of the circularization process and the disk formation of TDEs in the future.

\subsection{Accretion process in a forming disk}
\label{subsec_discuss_accretion}
In Section \ref{sec_fitting}, we point out that the scale-height ratio $H/R \simeq 0.1$, which is inferred from the short accretion timescale $t_{\rm acc} \sim 105$ days, is in conflict with the canonical value $H/R \simeq 0.01$ of a standard geometrically thin disk.

We would like to highlight that the empirical value of $H/R \simeq 0.01$ is based on the theory of a standard steady-state, radially extended accretion disk. In contrast, the TDE disk is transient and much more compact.

In fact, the disk is still receiving a continuous mass supply from the fallback stream tail. Due to the redistribution of the angular momentum during the circularization process and/or the Lense--Thirring effect, the angular momentum vector is misaligned with the disk orientation. Therefore, the debris will shock and heat up the disk when it supplies it, and thus the heating raises the disk height and accelerates the accretion process\footnote{It is similar to what \cite{Chan_Light_2020} found. They performed simulations to study the consequences of the interaction between the debris stream and the preexisting accretion disk. Although the disk we consider here is newly formed, the effects of the stream--disk interaction would be similar.}.

Furthermore, repeated encounters with shocks can remove angular momentum from the disk gas and cause it to move speedily inward. However, these issues need to be confirmed by the hydrodynamic simulation.

\subsection{Gravitational instability of the stream} \label{subsec_grav_instability}
In the calculation of the circularization process, we assume that the stream is collisional. However, if the matter recollapses into clumps in the stream due to gravitational instability when the self-gravity overcomes the tidal force and the pressure \citep{Coughlin_Variability_2015}, the subsequent self-crossing shock might be weaker, and the accretion of these clumps would affect the late light curve. Here we briefly discuss whether such effects would occur.

In the beginning of the disruption, the tidal force dominates in the stream. When the gas moves to a large distance from the central BH, the tidal force will become less important, and the gas will recollapse. In \cite{Coughlin_Variability_2015} performed a simulation of the tidal disruption of a Sun-like star by a $10^6\ M_{\odot}$ SMBH with $\beta=1$, and they found that the clumps fall back to the pericenter after $\sim 2$ yr since the disruption. For a partial TDE with total fallback mass of $\sim 0.02\ M_{\odot}$ that we obtain in the fitting of the AT 2019avd light curve, the self-gravity would be weaker due to the low density of the stream. In such case, even if the clumps could exist, their formation would be delayed, and thus the beginning of the clumps' fallback would be much later than $\sim 2$ yr. The duration of the flares in AT 2019avd is only $\sim 3$ yr; therefore, we would not see the consequences of gravitational instability in AT 2019avd. 

Furthermore, tidal force is not the only resistance for recollapse; we also need to consider the pressure of the stream. The Jeans condition $H \gtrsim c_{\rm s} \sqrt{\pi/(\rho G)}$ also needs to be satisfied, where $H$ and $c_{\rm s}$ are width of the stream and sound speed, respectively. Assuming a simple form of the stream density $\rho \sim \Delta M/ (\pi H^2 l_{\rm s})$, where $l_{\rm s}$ is the length of the stretched stream in the transverse direction, the Jeans condition becomes $c_{\rm s} \lesssim \sqrt{G \Delta M/(\pi l_{\rm s})}$. When the gas pressure dominates, the sound speed is $c_{\rm s} \sim \sqrt{k_{\rm b} T/(\mu m_{\rm p})}$, where $k_{\rm b}$, $\mu$, and $m_{\rm p}$ are the Boltzmann constant, mean particle weight, and photon mass, respectively. Then we find that the upper limit of the temperature for recollapse is
\begin{equation}
T \lesssim \frac{\mu m_{\rm p} G \Delta M}{\pi k_{\rm b} l_{\rm s}} \simeq 10 \mu \left(\frac{\Delta M}{0.02\ M_{\odot}} \right) \left( \frac{l_{\rm s}}{10^{15}\ {\rm cm}}\right)^{-1}\ {\rm K}. 
\label{eq:T_collapse}
\end{equation}
For a small amount of the fallback mass, $\sim 0.02\ M_{\odot}$, the gas would collapse only when the gas temperature is lower than $~10$ K. This condition could not be satisfied due to the heating by the cosmic rays and the radiation of the source. Therefore, we consider that recollapse by the gravitational instability should not occur in the partial TDE with a small amount of the fallback mass ($\lesssim 0.02\ M_{\odot}$).

\subsection{TDE Candidates with Double-peak Light Curves} \label{subsec_discuss_candidate}
There are other reported TDEs that have double-peak light curves, e.g., ASASSN-15lh and PS1-10adi. The former was first reported as a hydrogen-poor superluminous supernova \citep{Dong_15lh_2016}. Further monitoring indicates that ASASSN-15lh is more consistent with a TDE with a rapidly spinning and heavy SMBH ($\gtrsim 10^8 M_{\odot}$) than with a superluminous supernova \citep{Leloudas_15lh_2016,Kruhler_15lh_2018}. However, the likelihood of a TDE with such a high-mass SMBH is very low. The TDE PS1-10adi was reported as a TDE or a supernova occurring at the center of an active galaxy \citep{Kankare_10adi_2017}. The rebrightening of PS1-10adi in the optical/UV is very late, about $1500$ days after the main peak, with a short duration ($\sim 250$ days), and the infrared luminosity near the rebrightening is dimmer than that near the main peak. These features are different from AT 2019avd. It is considered that the rebrightening of PS1-10adi is powered by shock interaction between an outflow and a surrounding torus \citep{Jiang_Infrared_2019,Zhuang_Shock_2021}.

\section{Acknowledgments}
R.F.S. benefited tremendously from the discussion with Dr. Yanan Wang. This work is supported by National Natural Science Foundation of China (NSFC- 12073091, 11833007, 11733001), China Manned Spaced Project (CMS-CSST-2021-B11), Joint Research Foundation in Astronomy (U1731104, U2031106) under cooperative agreement between NSFC and CAS, and by Guangdong Basic and Applied Basic Research Foundation (2019A1515011119). This work made use of data supplied by the UK Swift Science Data Centre at the University of Leicesterm, the Near-Earth Object Wide-field Survey Explorer (NEOWISE), and the Zwicky Transient Facility (ZTF) project.

\software{emcee \citep{Foreman-Mackey_emcee_2013},
                HEAsoft (https://heasarc.gsfc.nasa.gov/lheasoft/)
               XSPEC \citep[V12.9, ][]{1996ASPC..101...17A}}


\begin{appendix}
\label{Appendix}
\section{Swift Follow-up and X-Ray spectral fit results}
All Swift XRT and UVOT follow-up observations are presented in Table~\ref{tab:xray} and Table~\ref{tab:uv}.


\begin{table*}
\centering
\caption{Log of Swift/XRT Observations and X-Ray Spectral Fit Results from the Extracted XRT Spectra.}
\label{tab:xray}
\begin{tabular}{lccccccc}
\toprule
MJD & Exposure & Count Rate & $kT$ & $F_{\rm 0.3-2\ keV}$ & $L_{\rm 0.3-2\ keV}$ & $L_{\rm bb}$ \\ & (ks) & ($10^{-3}\ {\rm cts\ s^{-1}}$) & (eV) & ($10^{-12}\ {\rm erg\ s^{-1}\ cm^{-2}}$) & ($10^{42}\ {\rm erg\ s^{-1}}$) & ($10^{42}\ {\rm erg\ s^{-1}}$) \\
\midrule
$58982$ & $1.61$ & $29.31 \pm 4.52$ & $66.7_{-12.6}^{+14.1}$ & $1.54 \pm 0.24$ & $3.16 \pm 0.49$ & $9.95 \pm 1.53$ \\
$58988$ & $0.49$ & $36.59 \pm 8.62$ & $104.0_{-30.8}^{+70.7}$ & $1.71 \pm 0.4$ & $3.4 \pm 0.8$ & $5.45 \pm 1.28$ \\
$58994$ & $1.73$ & $33.69 \pm 4.57$ & $122.0_{-15.6}^{+18.7}$ & $1.19 \pm 0.16$ & $2.34 \pm 0.32$ & $3.24 \pm 0.44$ \\
$59003$ & $1.98$ & $33.61 \pm 4.26$ & $106.0_{-14.5}^{+16.3}$ & $1.28 \pm 0.16$ & $2.53 \pm 0.32$ & $3.92 \pm 0.5$ \\
$59011$ & $1.96$ & $37.7 \pm 4.5$ & $93.7_{-11.3}^{+12.5}$ & $1.58 \pm 0.19$ & $3.15 \pm 0.38$ & $5.59 \pm 0.67$ \\
$59108$ & $2.95$ & $245.5 \pm 9.22$ & $115.0_{-4.0}^{+4.2}$ & $11.5 \pm 0.43$ & $22.72 \pm 0.85$ & $32.89 \pm 1.24$ \\
$59109$ & $0.28$ & $313.8 \pm 33.9$ & $101.0_{-10.3}^{+11.8}$ & $11.99 \pm 1.3$ & $23.84 \pm 2.58$ & $38.71 \pm 4.18$ \\
$59117$ & $1.03$ & $97.35 \pm 9.81$ & $106.0_{-9.9}^{+11.0}$ & $3.58 \pm 0.36$ & $7.09 \pm 0.71$ & $11.0 \pm 1.11$ \\
$59117$ & $0.88$ & $178.3 \pm 14.3$ & $90.4_{-8.2}^{+9.3}$ & $7.53 \pm 0.6$ & $15.08 \pm 1.21$ & $28.02 \pm 2.25$ \\
$59118$ & $0.35$ & $80.66 \pm 15.6$ & $101.0_{-18.8}^{+24.9}$ & $3.22 \pm 0.62$ & $6.4 \pm 1.24$ & $10.44 \pm 2.02$ \\
$59120$ & $0.85$ & $229.3 \pm 16.5$ & $114.0_{-7.6}^{+8.4}$ & $12.47 \pm 0.9$ & $24.65 \pm 1.77$ & $35.89 \pm 2.58$ \\
$59120$ & $0.63$ & $241.8 \pm 19.7$ & $109.0_{-9.1}^{+10.4}$ & $9.14 \pm 0.74$ & $18.11 \pm 1.48$ & $27.38 \pm 2.23$ \\
$59121$ & $0.47$ & $107.1 \pm 15.1$ & $109.0_{-15.5}^{+20.1}$ & $7.27 \pm 1.02$ & $14.39 \pm 2.03$ & $21.75 \pm 3.07$ \\
$59123$ & $0.82$ & $129.6 \pm 12.6$ & $103.0_{-9.2}^{+10.2}$ & $5.89 \pm 0.57$ & $11.71 \pm 1.14$ & $18.77 \pm 1.82$ \\
$59187$ & $0.95$ & $63.23 \pm 8.24$ & $103.0_{-14.6}^{+17.2}$ & $2.39 \pm 0.31$ & $4.75 \pm 0.62$ & $7.56 \pm 0.99$ \\
$59191$ & $0.52$ & $191.9 \pm 19.4$ & $110.0_{-10.2}^{+11.4}$ & $7.1 \pm 0.72$ & $14.06 \pm 1.42$ & $21.04 \pm 2.13$ \\
$59195$ & $0.27$ & $110.0 \pm 20.5$ & $102.0_{-18.8}^{+26.6}$ & $4.63 \pm 0.86$ & $9.2 \pm 1.71$ & $14.8 \pm 2.76$ \\
$59205$ & $0.75$ & $64.17 \pm 9.29$ & $85.6_{-12.0}^{+14.5}$ & $2.84 \pm 0.41$ & $5.7 \pm 0.83$ & $11.41 \pm 1.65$ \\
$59207$ & $0.94$ & $51.33 \pm 7.56$ & $99.0_{-14.3}^{+18.0}$ & $2.12 \pm 0.31$ & $4.23 \pm 0.62$ & $7.04 \pm 1.04$ \\
$59209$ & $1.02$ & $124.1 \pm 11.1$ & $96.5_{-8.3}^{+9.1}$ & $6.33 \pm 0.57$ & $12.62 \pm 1.13$ & $21.63 \pm 1.93$ \\
$59211$ & $0.97$ & $95.49 \pm 9.97$ & $97.3_{-10.1}^{+11.2}$ & $3.76 \pm 0.39$ & $7.49 \pm 0.78$ & $12.71 \pm 1.33$ \\
$59213$ & $0.92$ & $248.9 \pm 16.6$ & $110.0_{-6.7}^{+7.3}$ & $9.63 \pm 0.64$ & $19.07 \pm 1.27$ & $28.51 \pm 1.9$ \\
$59217$ & $0.56$ & $55.64 \pm 10.0$ & $123.0_{-19.2}^{+23.9}$ & $2.0 \pm 0.36$ & $3.94 \pm 0.71$ & $5.41 \pm 0.97$ \\
$59256$ & $1.42$ & $89.4 \pm 8.07$ & $102.0_{-10.9}^{+9.9}$ & $4.24 \pm 0.38$ & $8.43 \pm 0.76$ & $13.64 \pm 1.23$ \\
$59259$ & $1.19$ & $61.49 \pm 7.27$ & $107.0_{-11.3}^{+12.9}$ & $2.63 \pm 0.31$ & $5.2 \pm 0.61$ & $8.01 \pm 0.95$ \\
$59261$ & $1.35$ & $182.5 \pm 11.7$ & $112.0_{-6.9}^{+7.5}$ & $6.66 \pm 0.43$ & $13.18 \pm 0.84$ & $19.52 \pm 1.25$ \\
$59262$ & $1.63$ & $154.5 \pm 9.86$ & $96.9_{-6.1}^{+6.5}$ & $6.31 \pm 0.4$ & $12.59 \pm 0.8$ & $21.47 \pm 1.37$ \\
$59266$ & $1.21$ & $232.3 \pm 13.9$ & $119.0_{-7.2}^{+7.1}$ & $8.4 \pm 0.5$ & $16.56 \pm 0.99$ & $23.3 \pm 1.39$ \\
$59269$ & $0.18$ & $86.86 \pm 22.6$ & $67.6_{-20.8}^{+26.2}$ & $5.4 \pm 1.41$ & $11.1 \pm 2.89$ & $33.87 \pm 8.81$ \\
$59272$ & $1.59$ & $18.41 \pm 3.66$ & $72.2_{-15.9}^{+20.9}$ & $1.01 \pm 0.2$ & $2.06 \pm 0.41$ & $5.49 \pm 1.09$ \\
$59279$ & $0.76$ & $175.7 \pm 15.3$ & $107.0_{-9.3}^{+10.4}$ & $6.91 \pm 0.6$ & $13.7 \pm 1.19$ & $21.08 \pm 1.84$ \\
$59281$ & $0.25$ & $74.47 \pm 17.5$ & $86.3_{-19.5}^{+26.8}$ & $3.61 \pm 0.85$ & $7.31 \pm 1.72$ & $16.01 \pm 3.76$ \\
$59282$ & $0.55$ & $171.4 \pm 17.9$ & $92.1_{-9.1}^{+10.2}$ & $8.02 \pm 0.84$ & $16.04 \pm 1.68$ & $29.1 \pm 3.04$ \\
$59285$ & $1.53$ & $207.8 \pm 11.8$ & $111.0_{-7.0}^{+7.6}$ & $7.44 \pm 0.42$ & $14.72 \pm 0.84$ & $21.96 \pm 1.25$ \\
$59368$ & $1.67$ & $1.44 \pm 1.41$ & $38.8_{-29.3}^{+142.2}$ & $0.14 \pm 0.13$ & $0.31 \pm 0.3$ & $6.55 \pm 6.42$ \\
$59369$ & $2.31$ & $3.82 \pm 1.61$ & $49.2_{-23.7}^{+63.1}$ & $0.18 \pm 0.08$ & $0.39 \pm 0.17$ & $2.98 \pm 1.26$ \\
$59371$ & $6.45$ & $4.06 \pm 1.01$ & $136.5_{-44.0}^{+69.3}$ & $0.12 \pm 0.03$ & $0.24 \pm 0.06$ & $0.31 \pm 0.08$ \\
$59375$ & $2.78$ & $4.45 \pm 1.52$ & $126.9_{-37.7}^{+56.6}$ & $0.14 \pm 0.05$ & $0.27 \pm 0.09$ & $0.36 \pm 0.12$ \\
$59377$ & $1.80$ & $2.57 \pm 1.85$ & $115.6_{-58.1}^{+104.2}$ & $0.13 \pm 0.09$ & $0.25 \pm 0.18$ & $0.36 \pm 0.26$ \\
\bottomrule
\end{tabular}
\tablenotetext{}{$L_{\rm bb}$ is the single blackbody luminosity, and $kT$ is the corresponding temperature of blackbody. We set the equivalent Galactic neutral hydrogen column density in the direction of AT 2019avd to be $N_{\rm H} = 2.42 \times 10^{20}\ {\rm cm^{-2}}$ (see Section \ref{subsec:spec_Xray}).}
\end{table*}

\begin{table*}
\begin{minipage}{\textwidth}
\centering
\caption{Swift/UVOT Photometry.}
\label{tab:uv}
\begin{tabular}{cccccc}
\toprule
MJD & UVW1 & MJD & UVM2 & MJD & UVW2 \\
\midrule
$58,983$ & $18.29 \pm 0.05 (18.79)$ &$58,988$ & $18.55 \pm 0.11 (18.88)$ &$58,988$ & $18.53 \pm 0.1 (18.8)$ \\
$58,988$ & $18.52 \pm 0.17 (19.29)$ &$58,994$ & $18.52 \pm 0.07 (18.82)$ &$58,994$ & $18.57 \pm 0.06 (18.87)$ \\
$58,994$ & $18.11 \pm 0.07 (18.47)$ &$59,117$ & $18.64 \pm 0.09 (19.04)$ &$59,117$ & $18.59 \pm 0.07 (18.89)$ \\
$59,004$ & $18.14 \pm 0.04 (18.52)$ &$59,117$ & $18.54 \pm 0.09 (18.85)$ &$59,117$ & $18.54 \pm 0.08 (18.82)$ \\
$59,011$ & $18.05 \pm 0.04 (18.36)$ &$59,120$ & $18.48 \pm 0.09 (18.76)$ &$59,118$ & $18.46 \pm 0.07 (18.68)$ \\
$59,117$ & $18.06 \pm 0.07 (18.38)$ &$59,120$ & $18.5 \pm 0.09 (18.8)$ &$59,120$ & $18.53 \pm 0.08 (18.79)$ \\
$59,117$ & $17.98 \pm 0.07 (18.26)$ &$59,121$ & $18.55 \pm 0.12 (18.88)$ &$59,120$ & $18.61 \pm 0.09 (18.93)$ \\
$59,120$ & $18.06 \pm 0.07 (18.38)$ &$59,123$ & $18.52 \pm 0.09 (18.82)$ &$59,121$ & $18.58 \pm 0.1 (18.88)$ \\
$59,120$ & $18.07 \pm 0.09 (18.4)$ &$59,187$ & $18.53 \pm 0.08 (18.84)$ &$59,123$ & $18.64 \pm 0.08 (18.99)$ \\
$59,121$ & $18.17 \pm 0.1 (18.57)$ &$59,191$ & $18.63 \pm 0.11 (19.02)$ &$59,187$ & $18.68 \pm 0.07 (19.05)$ \\
$59,123$ & $18.06 \pm 0.07 (18.39)$ &$59,195$ & $18.52 \pm 0.15 (18.83)$ &$59,191$ & $18.59 \pm 0.07 (18.9)$ \\
$59,187$ & $18.05 \pm 0.06 (18.37)$ &$59,205$ & $18.68 \pm 0.1 (19.13)$ &$59,195$ & $18.52 \pm 0.13 (18.78)$ \\
$59,195$ & $17.96 \pm 0.11 (18.23)$ &$59,207$ & $18.48 \pm 0.08 (18.76)$ &$59,205$ & $18.77 \pm 0.09 (19.22)$ \\
$59,205$ & $18.12 \pm 0.07 (18.49)$ &$59,211$ & $18.44 \pm 0.08 (18.7)$ &$59,207$ & $18.65 \pm 0.07 (19.0)$ \\
$59,207$ & $18.06 \pm 0.07 (18.39)$ &$59,213$ & $18.52 \pm 0.08 (18.82)$ &$59,211$ & $18.61 \pm 0.07 (18.93)$ \\
$59,211$ & $17.98 \pm 0.06 (18.26)$ &$59,217$ & $18.56 \pm 0.1 (18.89)$ &$59,213$ & $18.48 \pm 0.07 (18.72)$ \\
$59,213$ & $18.02 \pm 0.06 (18.32)$ &$59,256$ & $18.62 \pm 0.07 (19.01)$ &$59,217$ & $18.59 \pm 0.07 (18.89)$ \\
$59,256$ & $18.23 \pm 0.08 (18.68)$ &$59,259$ & $18.59 \pm 0.07 (18.96)$ &$59,256$ & $18.69 \pm 0.07 (19.08)$ \\
$59,259$ & $18.19 \pm 0.09 (18.61)$ &$59,261$ & $18.7 \pm 0.07 (19.16)$ &$59,259$ & $18.65 \pm 0.07 (18.99)$ \\
$59,261$ & $18.18 \pm 0.08 (18.6)$ &$59,262$ & $18.88 \pm 0.07 (19.55)$ &$59,261$ & $18.78 \pm 0.07 (19.25)$ \\
$59,262$ & $18.32 \pm 0.08 (18.85)$ &$59,266$ & $18.66 \pm 0.07 (19.07)$ &$59,262$ & $18.98 \pm 0.08 (19.69)$ \\
$59,266$ & $18.38 \pm 0.09 (18.97)$ &$59,273$ & $18.77 \pm 0.08 (19.31)$ &$59,266$ & $18.73 \pm 0.08 (19.14)$ \\
$59,269$ & $18.25 \pm 0.09 (18.72)$ &$59,282$ & $18.87 \pm 0.12 (19.54)$ &$59,272$ & $18.83 \pm 0.07 (19.34)$ \\
$59,272$ & $18.28 \pm 0.07 (18.77)$ &$59,285$ & $18.74 \pm 0.07 (19.25)$ &$59,281$ & $18.93 \pm 0.17 (19.57)$ \\
$59,281$ & $18.42 \pm 0.12 (19.05)$ &$59,368$ & $19.16 \pm 0.1 (20.45)$ &$59,282$ & $18.68 \pm 0.11 (19.05)$ \\
$59,282$ & $18.35 \pm 0.1 (18.92)$ &$59,369$ & $19.05 \pm 0.09 (20.04)$ &$59,285$ & $18.77 \pm 0.07 (19.22)$ \\
$59,285$ & $18.38 \pm 0.08 (18.98)$ &$59,372$ & $19.21 \pm 0.07 (20.68)$ &$59,368$ & $19.34 \pm 0.09 (20.97)$ \\
$59,368$ & $18.74 \pm 0.11 (19.94)$ &$59,375$ & $19.21 \pm 0.09 (20.67)$ &$59,369$ & $19.27 \pm 0.08 (20.64)$ \\
$59,369$ & $18.81 \pm 0.08 (20.23)$ &$59,377$ & $19.09 \pm 0.1 (20.17)$ &$59,372$ & $19.24 \pm 0.05 (20.5)$ \\
$59,372$ & $18.77 \pm 0.07 (20.06)$ &$0$ & $0.0 \pm 0.0 (0.0)$ &$59,375$ & $19.37 \pm 0.07 (21.2)$ \\
$59,375$ & $18.68 \pm 0.09 (19.74)$ &$0$ & $0.0 \pm 0.0 (0.0)$ &$59,377$ & $19.32 \pm 0.09 (20.87)$ \\
$59,377$ & $18.82 \pm 0.13 (20.28)$ &$0$ & $0.0 \pm 0.0 (0.0)$ &$0$ & $0.0 \pm 0.0 (0.0)$ \\
\bottomrule
\end{tabular}
\end{minipage}
\tablenotetext{}{The corrected magnitudes are shown in the parenthesis (see Section \ref{subsec_swift_follow}).}
\end{table*}

\end{appendix}

\bibliography{cited}

\end{CJK*}
\end{document}